\documentclass{myar2e}

\pdfoutput=1

\usepackage{graphics,graphicx,aasdefs,ARAstroBib}

\newcommand{\Lstar}{$L^{*}$}

\newcommand{\hst}{\textit{HST}}

\newcommand{\chandra}{\textit{Chandra}}

\newcommand{\comptonlong}{\textit{Compton Gamma-Ray Observatory}}
\newcommand{\bepposax}{\textit{BeppoSAX}}

\newcommand{\swift}{\textit{Swift}}
\newcommand{\asca}{\textit{ASCA}}

\newcommand{\xmm}{\textit{XMM}}
\newcommand{\suzaku}{\textit{Suzaku}}

\newcommand{\hete}{\textit{HETE-2}}
\newcommand{\konus}{\textit{Konus-WIND}}

\newcommand{\fermi}{\textit{Fermi}}

\newcommand{\uJy}{\mbox{$\mu$Jy}}
\newcommand{\lya}{\mbox{Ly$\alpha$}}
\newcommand{\mnhi}{\mbox{$N_{H}$}}

\newcommand{\ergsec}{erg s$^{-1}$}

\newcommand{\cm}{{\rm cm}}

\newcommand{\Msun}{\mbox{$M_\odot$}}

\newcommand{\xray}{\mbox{X-ray}}

\def\simlt{\mathrel{\hbox{\rlap{\hbox{\lower4pt\hbox{$\sim$}}}\hbox{$<$}}}}
\def\simgt{\mathrel{\hbox{\rlap{\hbox{\lower4pt\hbox{$\sim$}}}\hbox{$>$}}}}

\makeatletter
\renewcommand{\section}{\@startsection{section}{1}{\z@}%
                                   {-3.5ex \@plus -1ex \@minus -.2ex}%
                                   {2.3ex \@plus.2ex}%
                                   {\normalfont\large\bfseries}}
\makeatother

\bibpunct[,]{(}{)}{;}{a}{}{,}

\begin{document}

\jname{Annu. Rev. Astron. Astrophys.}
\jyear{2009}
\jvol{1}

\title{Gamma-Ray Bursts in the \textit{\bfseries Swift} Era}

\author{N.~Gehrels,$^1$ E.~Ramirez-Ruiz,$^2$ and D.~B.~Fox$^3$
  \rule[-0.4cm]{0cm}{0.7cm}
  \affiliation{[1] NASA-Goddard Space Flight Center, Greenbelt,
    Maryland 20771; email: gehrels@milkyway.gsfc.nasa.gov
\vspace*{0.5\baselineskip}\newline
  [2] Department of Astronomy and Astrophysics, University of
  California, Santa Cruz, California 95064; email: enrico@ucolick.org
\vspace*{0.5\baselineskip}\newline
  [3] Department of Astronomy and Astrophysics, Pennsylvania State
  University, University Park, Pennsylvania 16802; email:
  dfox@astro.psu.edu}}

\begin{keywords}
  cosmology: early universe;
  galaxies: interstellar medium, high-redshift;
  gamma rays: observations, theory; 
  stars: Wolf-Rayet; neutrinos;
  supernovae: general; 
  gravitational waves
\end{keywords}

\begin{abstract}

With its rapid-response capability and multiwavelength complement of
instruments, the {\it Swift} satellite has transformed our physical
understanding of $\gamma$-ray bursts (GRBs).  Providing high-quality
observations of hundreds of bursts, and facilitating a wide range of
follow-up observations within seconds of each event, {\it Swift} has
revealed an unforeseen richness in observed burst properties, shed
light on the nature of short-duration bursts, and helped realize the
promise of GRBs as probes of the processes and environments of star
formation out to the earliest cosmic epochs.  These advances have
opened new perspectives on the nature and properties of burst central
engines, interactions with the burst environment from microparsec to
gigaparsec scales, and the possibilities for nonphotonic
signatures. Our understanding of these extreme cosmic sources has thus
advanced substantially; yet, more than 40 years after their
discovery, GRBs continue to present major challenges on both
observational and theoretical fronts.

\end{abstract}

\maketitle


\section{INTRODUCTION}
\label{sec1}

\subsection{Setting the Stage}

Gamma-ray bursts (GRBs) are among the most fascinating phenomena in
the Universe.  They are bright flashes of radiation with spectral
energy distributions peaking in the $\gamma$-ray band.  They have
durations measured in seconds and appear to be capable of producing
directed flows of relativistic matter with {\it kinetic} luminosities
exceeding $10^{53}$ \ergsec, making them the most luminous events
known.  All evidence points to a gravitational power source associated
with the cataclysmic formation of a relativistic star or to a
precursor stage whose inevitable end point is a stellar mass black
hole.

The field of GRB astronomy has been greatly stimulated by the launch
of the \swift\ satellite \citep{gehrels04} in 2004 with its rapid
response and panchromatic suite of instruments onboard, and by the
development of new-technology robotic telescopes on the ground.  A
multidisciplinary approach is now emerging with data combined across
the electromagnetic spectrum to learn about the physical processes at
play; ``spectral chauvinism'' can no longer be tolerated in the modern
study of GRBs.  Even non-photonic neutrino and gravitational wave
instruments are becoming more sensitive and may soon be detecting
signatures related to GRBs.

While interesting on their own, GRBs are now rapidly becoming powerful
tools to study detailed properties of the galaxies in which they are
embedded and of the universe in general.  Their apparent association
with massive star formation and their brilliant luminosities make them
unique probes of the high-redshift universe and galaxy evolution.
Absorption spectroscopy of GRB afterglows is being used to study the
ISM in evolving galaxies, complementary to the traditional studies of
quasar absorption line systems. Possibly the most interesting use of
GRBs in cosmology is as probes of the early phases of star and galaxy
formation, and the resulting reionization of the universe at $z \sim 6
- 20$.  GRBs are bright enough to be detectable, in principle, out to
much larger distances than those of the most luminous quasars or
galaxies detected at present. Thus, promptly localized GRBs could
serve as beacons which, shining through the pregalactic gas, provide
information about much earlier epochs in the history of the
universe. \\

Before the advent of \swift, the study of GRBs had evolved somewhat
unsystematically, and, as a result, the field has a great deal of
historical curiosities such as complex classifications schemes which
are now becoming streamlined as the field matures. Objects which were
once thought to be different are now found to be related and the style
of research has shifted from piecewise studies to a more general
statistical approach.  Although leaps in understanding can still come
from extraordinary events as we show in several examples in this
review, the applications to broader astrophysics are coming from the
compilations of hundreds of events. The literature on this subject has
therefore become quite large, and we apologize for referring now and
then only to the most recent comprehensive article in a given topic.
There are several recent summary articles that give excellent reviews
in specific areas related to GRBs.  These include the supernova-burst
connection \citep{wb06}, short GRBs \citep{nakar2007,lr07}, afterglows
\citep{vkw00,2007ChJAA_7_1Z} and theory \citep{meszaros02}. 
Our objective here is to summarize the field of GRB astronomy, 
from the \swift\ era and prior to the next steps with the
{\it Fermi Gamma Ray
 Observatory} \citep{atwood2009},
interpreting past findings while looking ahead to future capabilities and
potential
breakthroughs.

\subsection{A Burst of Progress}
The first sighting of a GRB came on July 2, 1967, from the military
{\it Vela} satellites monitoring for nuclear explosions in
verification of the Nuclear Test Ban Treaty \citep{vela}. These
$\gamma$-ray flashes, fortunately, proved to be different from the
man-made explosions that the satellites were designed to detect and a
new field of astrophysics was born.  Over the next 30 years, hundreds
of GRB detections were made. Frustratingly, they continued to vanish
too soon to get an accurate angular position for follow-up
observations. The reason for this is that $\gamma$-rays are
notoriously hard to focus, so $\gamma$-ray images are generally not
very sharp.

Before 1997, most of what we knew about GRBs was based on observations
from the Burst and Transient Source Experiment (BATSE) on board the
\comptonlong, whose results have been summarized in
\citet{batse}. BATSE, which measured about 3000 events, revealed that
between two and three visible bursts occur somewhere in the universe
on a typical day. While they are on, they can outshine every other
source in the $\gamma$-ray sky, including the Sun. Although each is
unique, the bursts fall into one of two rough categories. Bursts that
last less than two seconds are {\it short}, and those that last longer
-- the majority -- are {\it long}. The two categories differ
spectroscopically, with short bursts having relatively more
high-energy $\gamma$-rays than long bursts do.

Arguably the most important result from BATSE concerned the
distribution of bursts. They occur isotropically -- that is, evenly
over the entire sky $-$ suggesting a cosmological distribution with no
dipole and quadrupole components. This finding cast doubt on the
prevailing wisdom, which held that bursts came from sources within the
Milky Way.  Unfortunately, $\gamma$-rays alone did not provide enough
information to settle the question for sure. The detection of
radiation from bursts at other wavelengths would turn out to be
essential. Visible light, for example, could reveal the galaxies in
which the bursts took place, allowing their distances to be
measured. Attempts were made to detect these burst counterparts, but
they proved fruitless.

A watershed event occurred in 1997, when the \bepposax\ satellite
succeeded in obtaining high-resolution X-ray images \citep{beppo} of
the predicted fading afterglow of GRB 970228 -- so named because it
occurred on February 28, 1997. This detection, followed by a number of
others at an approximate rate of 10 per year, led to positions
accurate to about an arc minute, which allowed the detection and
follow-up of the afterglows at optical and longer
wavelengths\footnote{We note, however, that the first optical
  afterglow detection of GRB 970228 \citep{vgg97}
was based on the        X-ray prompt
  detection by \bepposax.} 
  (e.g., \citealt{vgg97}).
    This paved the way for the measurement of redshift
distances, the identification of candidate host galaxies, and the
confirmation that they were at cosmological distances \citep{firstz}.

Among the first GRBs pinpointed by \bepposax was GRB 970508
\citep{firstz}.  Radio observations of its afterglow provided an
essential clue. The glow varied erratically by roughly a factor of two
during the first three weeks, after which it stabilised and then began
to diminish \citep{1997Natur_389_261F}.  The large variations probably
had nothing to do with the burst source itself; rather they involved
the propagation of the afterglow light through space. Just as the
Earth's atmosphere causes visible starlight to twinkle, interstellar
plasma causes radio waves to scintillate. Therefore, if GRB 970508 was
scintillating at radio wavelengths and then stopped, its source must
have grown from a mere point to a discernible disk. ``Discernible''
here means a few light-weeks across. To reach this size the source
must have been expanding at a considerable rate -- close to the speed
of light \citep{waxman98}.

The observational basis for a connection between GRBs and supernovae
was prompted by the discovery that the {\it BeppoSAX} error box of GRB
980425 contained supernova SN 1998bw \citep{1998Natur_395_670G}.  A
number of other GRBs have since shown a 1998bw-like temporal component
superposed on the power law optical lightcurve
\citep{wb06}, but they still lacked a clear spectroscopic detection of
an underlying supernova. Detection of such signature came with the
discovery by the HETE-2 mission of GRB 030329
(\citealt{2003ApJ_591L_17S}, \citealt{Hjorth:2003:VES}).  Due to its
extreme brightness and slow decay, spectroscopic observations were
extensive. The early spectra consisted of a power-law decay continuum
($F_\nu \propto \nu^{-0.9}$) typical of GRB afterglows with narrow
emission features identifiable as H$\alpha$, [OIII], H$\beta$ and
[OII] at $z=0.1687$ (\citealt{2003ApJ_593L_19K},
\citealt{2003ApJ_599_394M}) making GRB 030329 the second nearest burst
overall at the time and the nearest {\it classical} burst\footnote{The
  other GRBs with supernova associations have been underluminous
  events \citep{2007ApJ_654_385K}.}. A major contribution to our
understanding of X-ray prompt emission came also from the HETE-2
mission \citep{2004NewAR_48_423L} which was active from 2000 to 2006.
Dozens of bursts in the ``X-ray flash'' category were observed and
were found to be similar in origin to the classical long GRBs
\citep{2004BaltA_13_201M}.

\swift\ is the current GRB discovery mission.  It is a space robot
designed specifically with GRBs in mind.  It combines a wide-field
hard X-ray burst detection telescope (Burst Alert Telescope - BAT;
\citealt{bbc+05}), with narrow-field X-ray (X-Ray Telescope - XRT;
\citealt{bhn+05}), and UV-optical (UV Optical Telescope - UVOT;
\citealt{rhm+04}) telescopes.  A powerful and fast on-board burst
detection algorithm \citep{2003AIPC_662_491F} provides the burst
coordinates to the spacecraft, which autonomously repoints the
observatory so that X-ray and optical observations typically commence
within two minutes of the burst trigger.  The mission was designed to
find counterparts for all burst types, including the
previously-elusive short GRBs.  Burst positions and other data are
provided promptly to the ground for ground observers.

The burst detection rate for \swift\ is about 100 GRBs per year,
resulting in a current data set as of December 2008 of 380 bursts.  Of
these, there are 126 with redshift determination, mostly from
spectrographs on large optical telescopes and new robotic telescopes
on the ground.  These now far outnumber the $\sim40$ GRB redshifts
available prior to \swift.  More than 95\% of the \swift\ bursts have
X-ray afterglow detection and $\sim60$\% have optical afterglows (UVOT
+ ground).  To date 33 {\it short} bursts have been localized with 8
having redshift determinations. The new data have enabled much more
detailed studies of the burst environment, the host galaxy, and the
intergalactic medium between galaxies.  \swift\ and follow-up
observations have also transformed our view of GRB sources. For
example, as discussed in Section \ref{sec3}, the old concept of a
sudden release of energy concentrated in a few brief seconds has been
discarded. Indeed, even the term ``afterglow'' is now recognized as
misleading -- the energy radiated during both phases is comparable.

Our primary intention in this review is to describe the most important
observational discoveries in the \swift\ era as well as to explain
some of the new understanding of key mechanisms that are believed to
operate in these objects to an astronomical audience with little prior
exposure to GRBs. Four sections follow.  Section \ref{sec2} provides a
description of our current knowledge of what constitutes a GRB.
Section \ref{sec3} is a summary of the observations of the prompt and
afterglow emission and Section IV of the observations of host galaxies
and progenitor clues.  In Section V, we examine our current progress
in our understanding of the basic physical processes at work.  Section
VI is a look forward at future prospect for GRB study.



\section{WHAT IS A GAMMA-RAY BURST?}
\label{sec2}

\begin{figure}
\begin{center}
\includegraphics[scale=0.6]{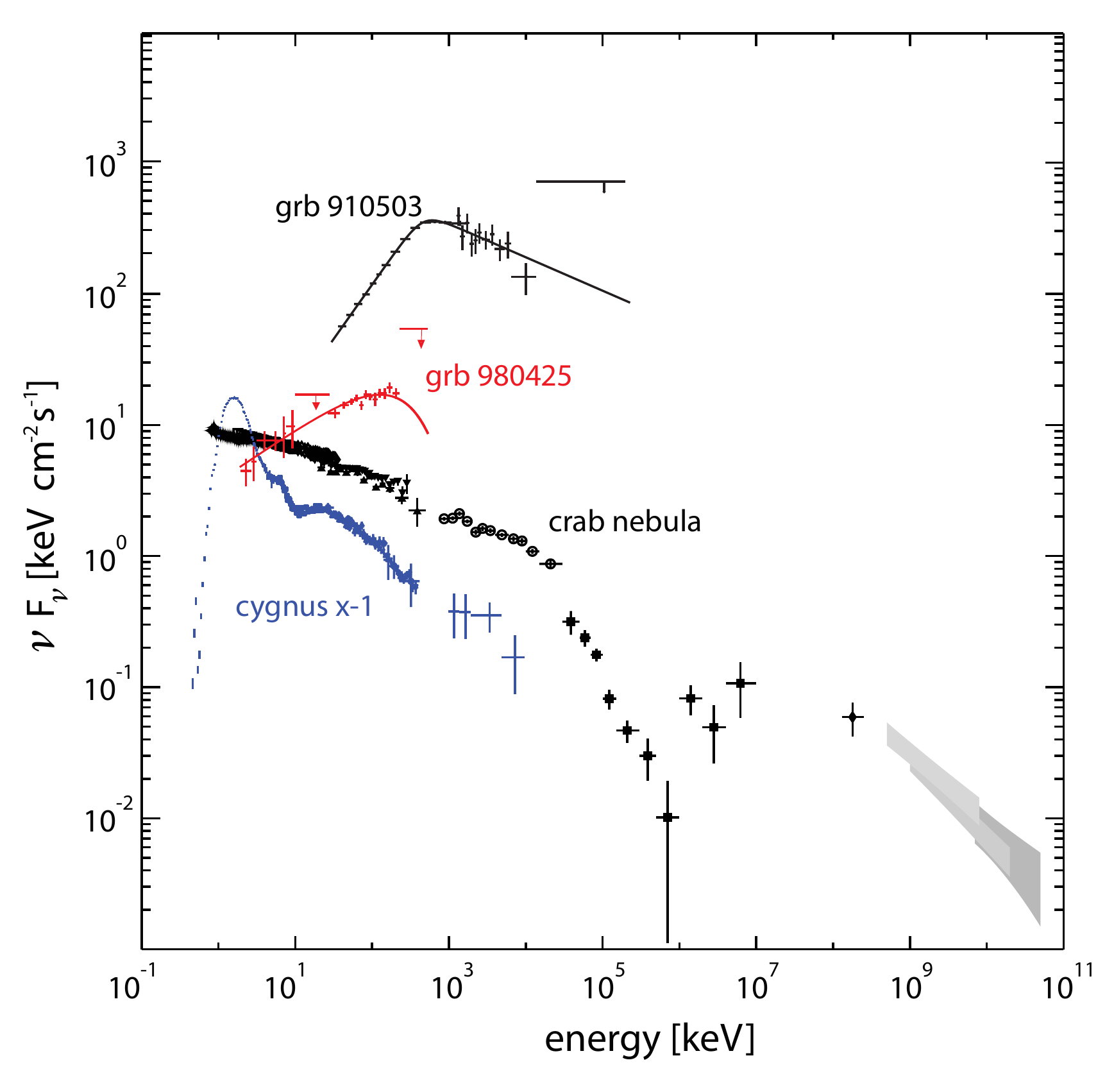}
\end{center}
\caption{\small
  Gamma-rays are excellent probes of the most energetic phenomena in
  nature, that typically involve dynamical non-thermal processes and
  include interactions of high energy electrons with matter, photons
  and magnetic fields; high energy nuclear interactions; matter-
  antimatter annihilation and possibly other fundamental particle
  interactions. Here shown are representative spectra $\nu F_{\nu}
  \propto \nu^2 N(\nu)$ of GRBs (\citealt{2007ApJ_654_385K};
  \citealt{2008ApJ_677_1168K}) together with the Crab pulsar nebula
  \citep{2001AA_378_918K} and the galactic black hole candidate Cygnus
  X-1 \citep{2002ApJ_572_984M}.}
\label{figIIa}
\end{figure}

GRBs are sudden, intense flashes of $\gamma-$rays which, for a few
blinding seconds, light up an otherwise fairly faint $\gamma-$ray sky.
Spectra extending over many decades in photon energy have now been
measured for hundreds of GRBs. In Figure~\ref{figIIa} representative
spectra are plotted in the conventional coordinates $\nu$ and $\nu
F_\nu$, the energy radiated per logarithmic (natural log) frequency
interval. Some obvious points should be emphasized. First, we measure
directly only the energy radiated in the direction of the earth per
second per steradian per logarithmic frequency interval by a
source. The apparent bolometric luminosity may be quite different from
the {\it true} bolometric luminosity if the source is not emitting
isotropically.  Second, there is striking evidence for a
characteristic photon energy (peak in the $\nu F_\nu$ spectrum), which
appears to be related to the overall spectral luminosity
normalization. In contrast, the spectra of many galactic and
extragalactic accretion systems are often well fitted by single
power-laws. A simple power-law contains little information, whereas a
complex spectrum composed of many broken power-laws tell us much more,
as each break frequency must be explained.

\begin{figure}       
\begin{center}
\includegraphics[scale=0.7]{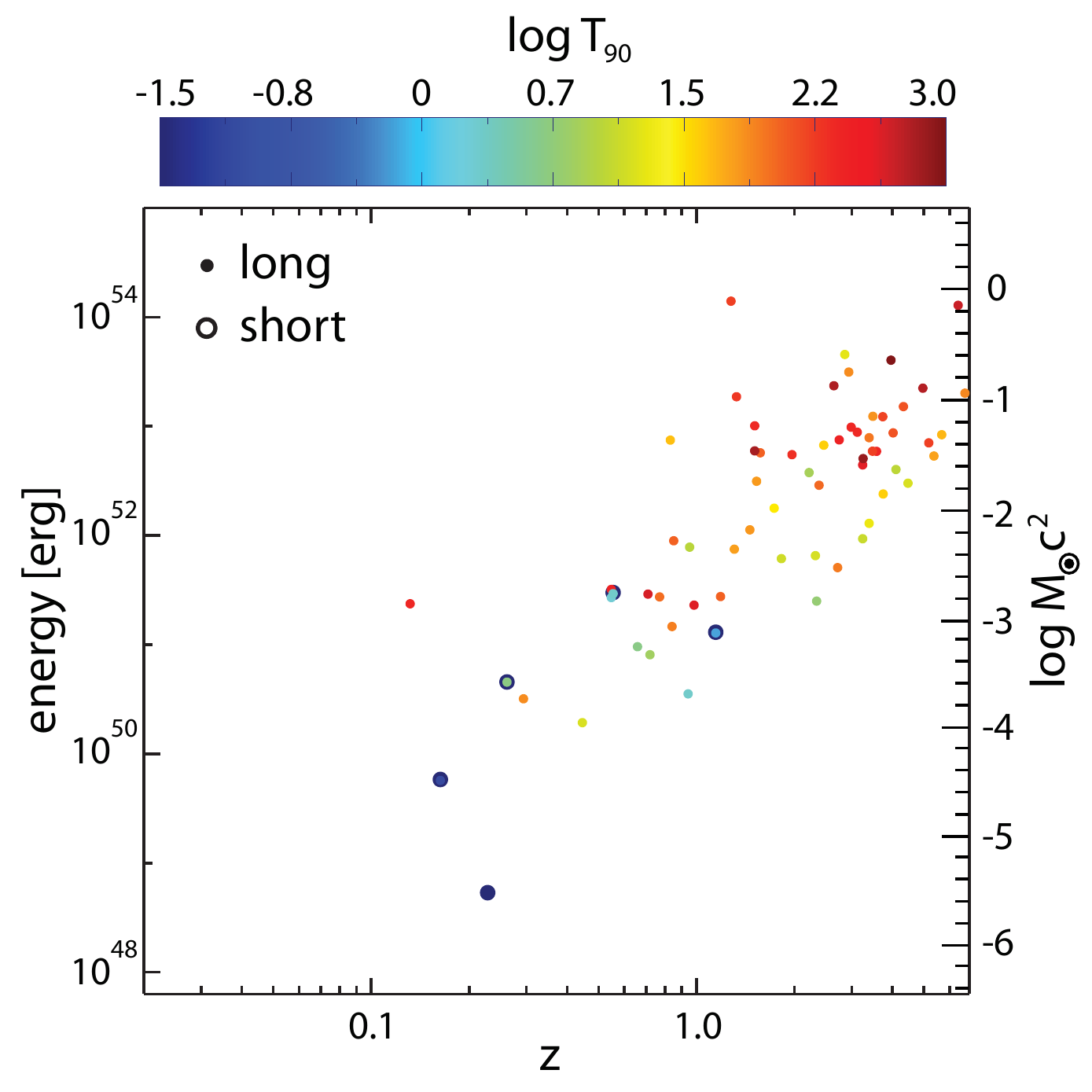}
\end{center}
\caption{\small
  Apparent isotropic $\gamma$-ray energy as a function
  of redshift and observed duration. The energy is calculated assuming
  isotropic emission in a common comoving bandpass for a sample of
  short and long GRBs with measured redshifts. This spread in the
  inferred luminosities obtained under the assumption of isotropic
  emission may be reduced if most GRB outflows are jet-like. A beamed
  jet would alleviate the energy requirements, and some observational
  evidence does suggest the presence of a jet.}
\label{figIIb}
\end{figure}

At cosmological distances the observed GRB fluxes imply energies that
can exceed $10^{53}(\Omega/4\pi)$ erg, where $\Omega$ is the solid
angle of the emitting region (Figure~\ref{figIIb}, see also
\citealt{2001AJ_121_2879B}).  This is the mass equivalent of
$0.06\Msun$ for the isotropic case. Compared with the size of the sun,
the seat of this activity is extraordinarily compact, with sizes of
less than milli-light-seconds ($<$300 km) as indicated by rapid
variability of the radiation flux \citep{1992Natur_359_217B}.  It is
unlikely that mass can be converted into energy with better than a few
(up to ten) percent efficiency; therefore, the more powerful GRB
sources must ``process'' upwards of $10^{-1}(\Omega/4\pi)M_\odot$
through a region which is not much larger than the size of a neutron
star (NS) or a stellar mass black hole (BH). No other entity can
convert mass to energy with such a high efficiency, or within such a
small volume.

The observed $\gamma$-rays have a nonthermal spectrum. Moreover, they
commonly extend to energies above 1 MeV, the pair production threshold
in the rest frame.  These facts together imply that the emitting
region must be relativistically expanding
(\citealt{1983MNRAS_205_593G}, \citealt{1986ApJ_308L_47G},
\citealt{1986ApJ_308L_43P}).  We draw this conclusion for two
reasons. First, if the region were indeed only a light second across
or less, as would be implied by the observed rapid variability in the
absence of relativistic effects, the total mass of baryons in the
region would need to be below about $10^{-12}M_\odot$ in order that
the electrons associated with the baryons should not provide a large
opacity (\citealt{1993ApJ_403L_67P}, \citealt{1990ApJ_363_218P}).
Second, larger source dimensions are required in order to avoid
opacity due to photon-photon collisions.  If the emitting region is
expanding relativistically, then, for a given observed variation
timescale, the dimension $R$ can be increased by $\Gamma^2$. The
opacity to electrons and pairs is then reduced by $\Gamma^4$, and the
threshold for pair production, in the observer frame, goes up by 
$\sim\Gamma$
 from its rest frame value (\citealt{1993AAS_97_59F},
\citealt{1995ApJ_453_583W}, \citealt{1997ApJ_491_663B},
\citealt{2008ApJ_677_92G}).  Best-guess numbers are Lorentz factors
$\Gamma$ in the range $10^2$ to $10^3$ \citep{2001ApJ_555_540L},
allowing rapidly-variable emission to occur at radii in the range
$10^{12}$ to $10^{14}$ cm.

\smallskip

\begin{figure}
\begin{center}
\includegraphics[width=5.5in]{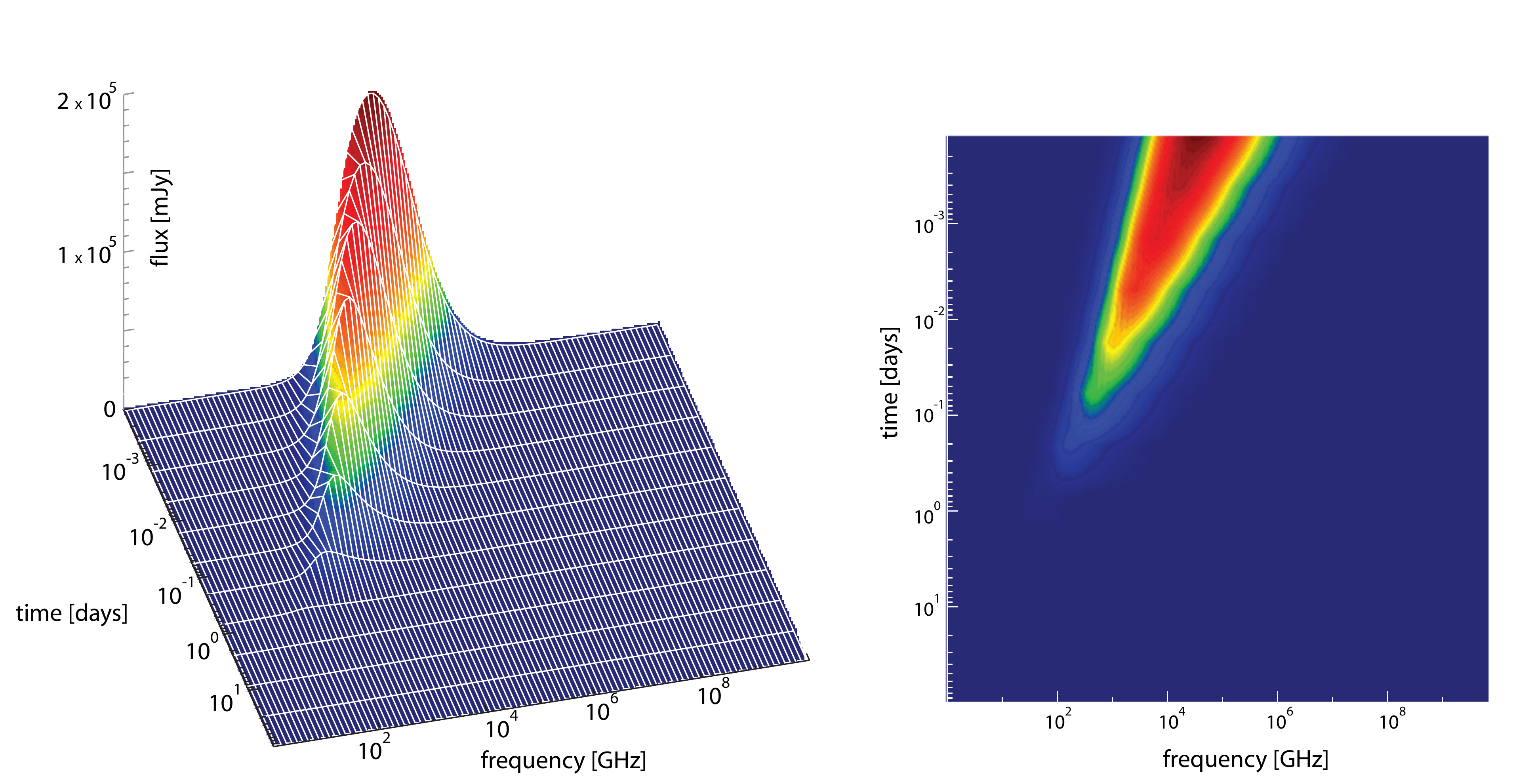}
\end{center}
\caption{\small
   The evolving synchrotron afterglow of a $\gamma-$ray
  burst.  Shown is a theoretical model \citep{gfm07} for the afterglow
  of the {\it Swift} GRB\,050904.  The model is presented without
  extinction and as it would have been observed at redshift $z=2$; the
  burst itself occurred at $z=6.29$.  The evolution of the synchrotron
  peak to lower frequencies is clearly visible. More subtle effects
  including evolution of the synchrotron cooling and self-absorption
  frequencies, and the associated synchrotron self-Compton emission of
  the blastwave at higher frequencies, are not readily visible in this
  model.}
\label{figIIc}
\end{figure}

Because the emitting region must be several powers of ten larger than
the compact object that acts as a trigger, there are further physical
requirements.  The original internal energy contained in the radiation
and pairs would, after expansion, be transformed into relativistic
kinetic energy.  A variant that has also been suggested is based on
the possibility that a fraction of the energy is carried by Poynting
flux (\citealt{Blandford:1977:EEE},
\citealt{1992Natur_357_472U}). This energy cannot be efficiently
radiated as $\gamma$-rays unless it is re-randomized
(\citealt{1994ApJ_432_181M}, \citealt{1992ApJ_395L_83N},
\citealt{1994ApJ_430L_93R}, \citealt{1994ApJ_427_708P}).  Impact on an
external medium (or an intense external radiation field; e.g.,
\citealt{1995MNRAS.277..287S}) would randomize half of the initial
energy merely by reducing the expansion Lorentz factor by a factor of
2.  For an approximately smooth distribution of external matter, the
bulk Lorentz factor of the fireball thereafter decreases as an inverse
power of the time. In the presence of turbulent magnetic fields built
up behind the shocks \citep{1992MNRAS_258P_41R}, the electrons produce
a synchrotron power-law radiation spectrum which softens in time, as
the synchrotron peak corresponding to the minimum Lorentz factor and
field decreases during the deceleration (\citealt{1994ApJ_422_248K},
\citealt{1996ApJ_473_204S}).  Thus, the GRB radiation, which started
out concentrated in the $\gamma$-ray range during the burst, is
expected to progressively evolve into an afterglow radiation which
peaks in the X-rays, then UV, optical, IR and radio (Figure
\ref{figIIc}).  Detailed predictions \citep{1997ApJ_476_232M} of the
afterglow properties, made in advance of the observations, agreed well
with subsequent detections at these photon energies, followed up over
periods of up to months.  The detection of diffractive scintillation
in the radio afterglow of GRB 970508, provided the first determination
of the source size and a direct confirmation of relativistic source
expansion \citep{1997Natur_389_261F}, which was further strengthened
by the size measurement of the afterglow image of GRB 030329 by radio
interferometry with the VLBA \citep{2004ApJ_609L_1T}.

The complex time-structure of some bursts suggests that the central
engine may remain active for up to 100 s \citep{2000ApJ_539_712R} or
possibly longer \citep{2007ApJ_671_1921F}.  However, at much later
times all memory of the initial time-structure would be lost:
essentially all that matters is how much energy and momentum has been
injected, its distribution in angle and velocity.  We can at present
only infer the energy per solid angle, but there are reasons to
suspect that bursts are far from isotropic. Due to relativistic
beaming, an observer will receive most emission from those portions of
a GRB blast wave that are within an angle $\sim 1/\Gamma$ of the
direction to the observer.  The afterglow is thus a signature of the
geometry of the ejecta - at late stages, if the outflow is beamed, we
expect a spherically-symmetric assumption to be inadequate; the
deviations from the predictions of such a model would then tell us
about the ejection in directions away from our line of sight
\citep{Rhoads:1999:DLC}.

The appearance of achromatic breaks in the development of GRB
afterglows has been interpreted as indicating that they are jet flows
beamed towards us. Collimation factors of $\Omega_i/4\pi \lesssim
0.01$ (corresponding to half opening angles of $\lesssim$ 8 degrees)
have been derived from such steepening (\citealt{fks+01},
\citealt{bloom03}).  If GRB sources are beamed, then this reduces the
energy per burst by two or three orders of magnitude at the expense of
increasing their overall frequency.

As regards the central engine trigger, there remain a number of key
questions. What are the progenitors? What is the nature of the
triggering mechanism, the transport of the energy and the time scales
involved?  Does it involve a hyperaccreting compact object? If so, can
we decide between the various alternative ways of forming it?  The
presence of some GRBs (in the short duration category) in old stellar
populations rules out a source uniquely associated with recent star
formation and, in particular, massive star origin for all bursts
(\citealt{2005Natur_437_851G}, \citealt{2006ApJ_638_354B}).  An
understanding of the nature of these sources is thus inextricably
linked to the {\it metabolic pathways} through which gravity, spin,
and energy can combine to form collimated, ultrarelativistic
outflows. These threads are few and fragile, and the tapestry is as
yet a poor image of the real universe. If we are to improve our
picture-making we must make more and stronger ties to physical theory.
But in reconstructing the creature, we must be guided by our eyes and
their extensions.  The following sections provide a detail summary of
the observed properties of these ultra-energetic phenomena.  These
threads will be woven in Section \ref{sec:theory}.

\section{BURST AND AFTERGLOW OBSERVATIONS}
\label{sec3}

The most direct diagnostics of the conditions within GRBs come from
the radiation we observe, which we summarize in this section.  There
are many ways to organize a discussion of the properties of the
radiation emanating from GRB sources.  We do not intend in this
article to a give a detailed review of individual events since there
are now sufficiently many examples that we are likely to be led
seriously astray if we test our theories against individual events.
For this reason, we center our discussion on major trends even in
cases in which the generalizations we describe are based on data that
do not yet have the statistical weight of a "complete" sample.  It
should be noted that there are inherent biases in the discovery of a
GRB at a given redshift which are often difficult to quantify, such as
complex trigger efficiencies and non-detections. Continued advances in
the observations will surely yield unexpected revisions and additions
in our understanding of the properties of GRBs.

\subsection{Prompt High-Energy Emission}
\subsubsection{Taxonomy.}
The manifestations of GRB activity are extremely diverse. GRBs are
observed throughout the electromagnetic spectrum, from GHz radio waves
to GeV $\gamma$-rays, but until recently, they were known
predominantly as bursts of $\gamma$-rays, largely devoid of any
observable traces at any other wavelengths.  $\gamma$-ray properties
provide only one of several criteria for classifying GRB sources.
Part of the problem is observational because it is not possible to
obtain full spectral coverage in all objects and it is not easy to
reconcile a classification based on host galaxy properties with one
based on the prompt $\gamma$-ray properties. The major impediment to
serious taxonomy is more fundamental. GRBs are heterogeneous objects,
especially in their directly observed properties.  The success of a
classification scheme, we believe, should be measured by the extent to
which newly recognized properties distinguish subsets defined by
differences in other properties. By this criterion, the taxonomy of
GRBs has met with only mixed success. As new non-$\gamma$-ray
selection techniques are introduced (e.g. age of stellar populations
in host galaxies or the presence of type Ic supernova signatures), the
class boundaries (e.g. short and long duration events) have blurred
where the defined subclasses transcend traditional boundaries. On the
other hand, many new properties do correlate with old ones. This is
all the more remarkable in that the conventional diagnostics
(e.g. burst duration) measure properties on scales several orders of
magnitude larger than that which we believe to be the characteristic
of the powerhouse.

\subsubsection{Observed durations and redshifts.}
GRB traditionally have been assigned to different classes based on
their duration - usually defined by the time during which the middle
50\% ($T_{50}$) or 90\% ($T_{90}$) of the counts above background are
measured.  On the basis of this criterion, there are two classes of
GRBs - short and long - where $\sim$2 s duration separates them.  The
initial hints for the existence of such classes
\citep{cline74,mazets81}, were followed by stronger evidence from
ISEE-3 and \konus\ data \citep{norris84} and definite proof using
large statistics from BATSE \citep{kouvel93}. BATSE results also
showed that short bursts have a harder spectrum than long bursts
\citep{kouvel93} although this is less prominent in observations by
\konus, \hete, and \swift\ \citep{sakamoto06}.

\begin{figure}
\begin{center}
\includegraphics[width=5.5in]{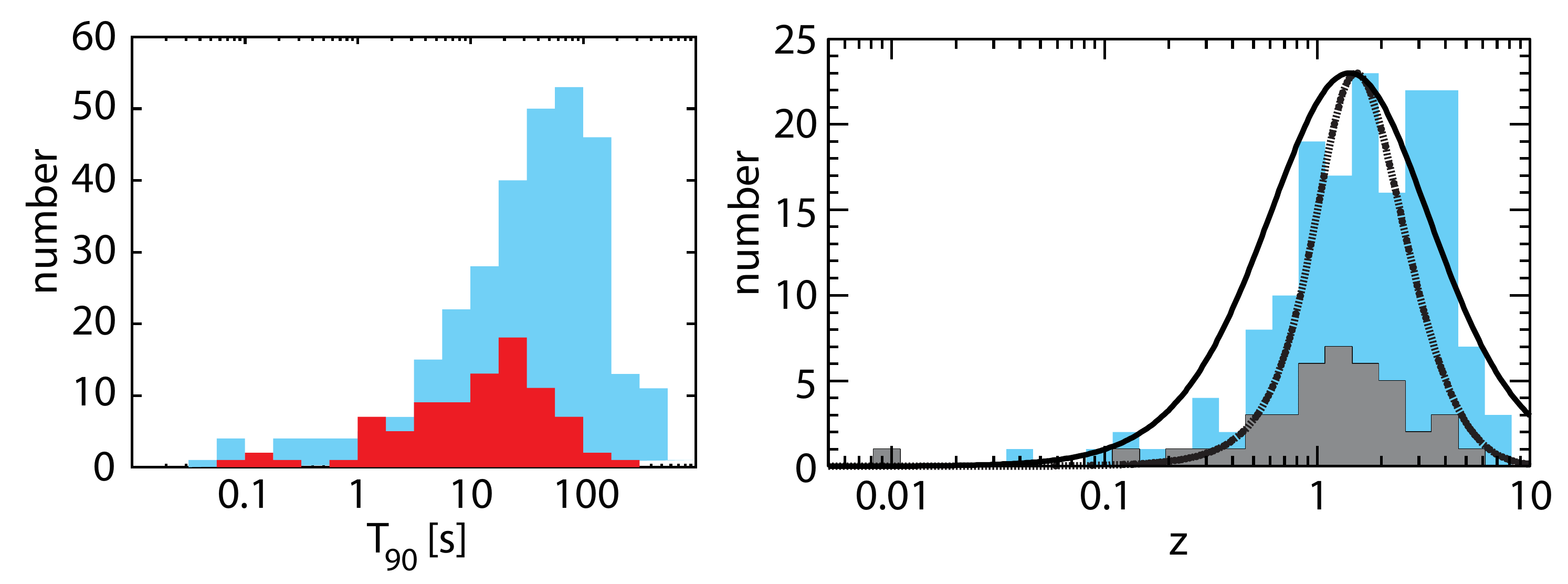}
\end{center}
\caption{\small
   Duration and redshift distributions for
  \swift\ GRBs. {\it Left} panel shows the duration distribution.  The
  blue histogram is the measured $T_{90}$ distribution while the
  orange one is corrected to the source frame: $T_{90} / (1+z)$. {\it
    Right} panel shows the redshift distribution for \swift\ GRBs in
  blue and pre-\swift\ GRBs in gray. \swift\ is detecting higher
  redshift bursts on average than pre-\swift.  The solid (broad)
  theory curve illustrates the evolution of a comoving volume element
  of the universe while the dotted (narrow) curve is a convolution of
  the comoving volume with a model for the star formation rate as
  calculated by \citet{2001ApJ_548_522P}.  }
\label{figIIIa}
\end{figure}

The duration and redshift distributions for \swift\ GRBs are shown in
Figure \ref{figIIIa}. The blue histogram in the {\it left} panel is
for observed durations while the orange histogram shows the durations
corrected to the source frame $T_{90}/(1+z)$ for those bursts with
redshift determinations.  In the source frame, the typical long burst
duration is $\sim20$s compared to $\sim50$s in the observer frame.
\swift\ has been detecting a lower fraction ($\sim$10\%) of short
bursts than BATSE did (25\%).  This is because \swift\ observes in a
softer energy band ($15-150$ keV) than BATSE (50 keV $-$ 2 MeV) and
because \swift\ requires a sky image of the event for burst detection
and the image part of the trigger algorithm is less sensitive to short
bursts due to their lower fluences.  Shown in Figure \ref{figIIIa}
({\it right} panel) are the measured redshift distributions. The blue
histogram is for \swift\ events while the gray one is for
pre-\swift\ bursts.  As clearly seen, \swift\ is currently detecting
GRBs at a higher average redshift: $<z>\sim 2.5$ for \swift\ bursts
while $<z>\sim 1.2$ for pre-\swift\ events.  The reason for this
difference is the higher sensitivity of \swift\ compared to
\bepposax\ and \hete.

GRBs have also been classified according to their spectral properties,
albeit less successfully.  In particular, bursts with lower spectral
energy peaks ($E_{\rm peak}$) have been denoted X-Ray Flashes (XRFs)
based on observations by \bepposax, BATSE and
\hete\ (\citealt{heise01}, \citealt{barraud03}, \citealt{kippen03},
\citealt{sakamoto05}).  These events are closely related to common
long duration GRBs and appear to form a continuum of all parameters
between the two types with no striking evidence for a discerning
characteristic \citep{2005ApJ_630_1003G}.

\subsubsection{Observed correlations.}
\begin{figure}[p!]
\begin{center}
\includegraphics[scale=0.6]{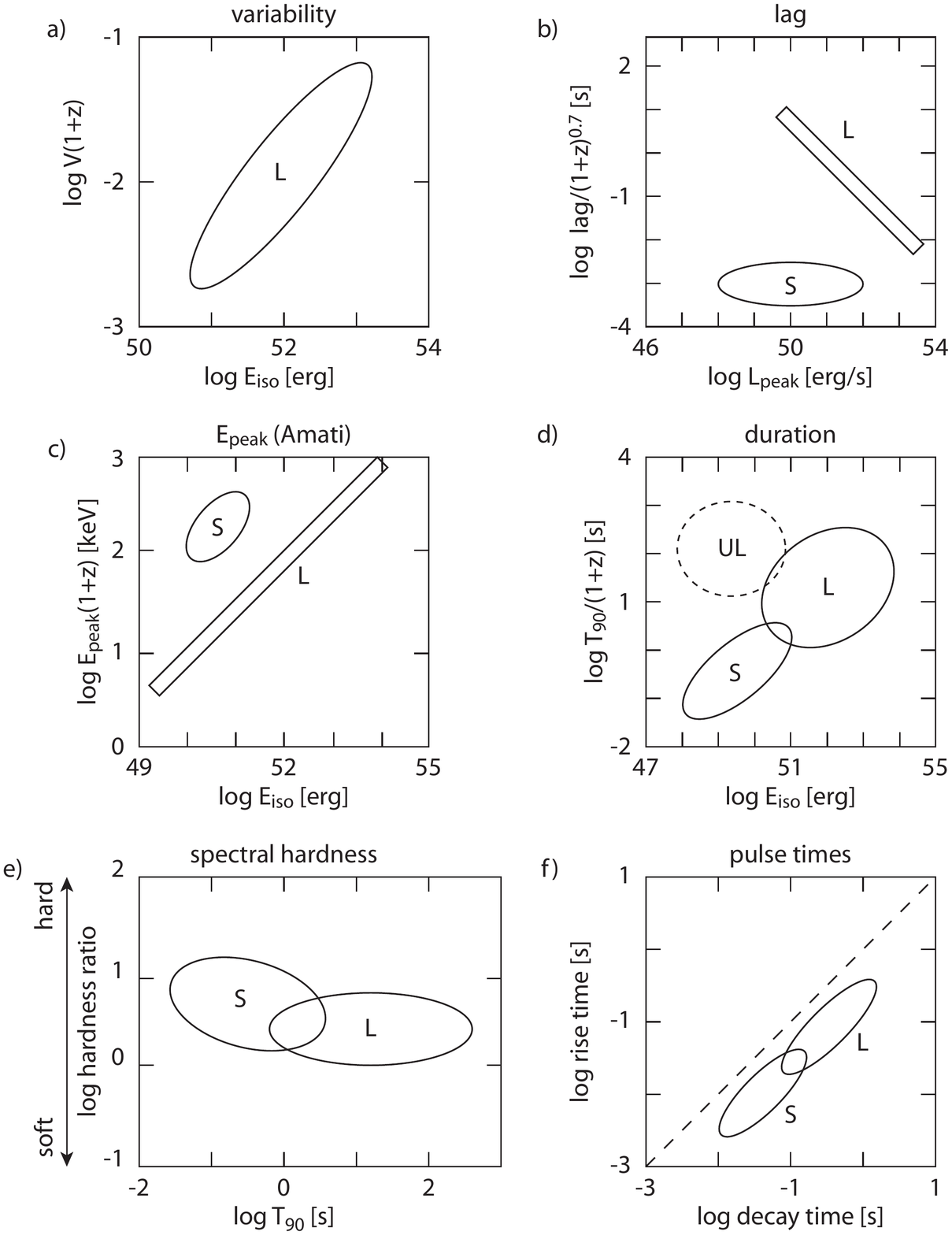}
\end{center}
\caption{\small\setlength{\baselineskip}{0.6\baselineskip}%
  Schematic diagrams illustrating the most widely
  discussed correlations between various prompt emission properties
  for long (L), short (S), and underluminous (UL) GRBs. (a)
  Variability scaled to the burst frame versus $E_{\rm iso}$
  \citep{fenimore00,2001ApJ...552...57R,schaefer06}. The variability
  is a measure of the spikiness of the lightcurve and is defined as
  the mean square of the time signal after removing low frequencies by
  smoothing.  (b) Spectral lag scaled to the burst frame versus peak
  luminosity (\citealt{norris_bonnell06}; \citealt{gehrels06}).  (c)
  $E_{\rm peak}$ scaled to the burst frame versus $E_{\rm iso}$
  (\citealt{aft+02} for \bepposax\ GRBs; \citealt{amati06} for
  \swift\ GRBs; \citealt{2002ApJ_576_101L} for BATSE events). (d)
  Duration scaled to the burst frame versus $E_{\rm iso}$. (e)
  Spectral hardness versus observed duration \citep{kouvel93}. (f)
  Pulse rise time versus its decay time \citep{norris96}.  }
\label{figIIIb}
\end{figure}

There is a great deal of diversity in the $\gamma$-ray prompt
lightcurves of GRBs.  Both long and short bursts can have temporal
profiles ranging from smooth, single-peaked pulses to highly structured 
multi-pulses.  The prompt emission can be characterized by a variety
of spectral and temporal parameters which include duration,
variability, lag, pulse rise/fall time, fluence, $E_{\rm iso}$ and
$E_{\rm peak}$. A schematic diagram illustrating the most widely
discussed $\gamma$-ray prompt correlations is shown in Figure
\ref{figIIIb} with detailed references given in its caption.  These
correlations are often based upon statistical analysis of quantities
whose physical causes are poorly understood, but almost certainly
depend on many variables. Interpretations of these correlations must
therefore be done with caution.

The prompt GRB lightcurves can generally be dissembled into a
superposition of individual pulses as described by \citet{norris96}
with rise times shorter on average than decay times ({\it panel f}).
The variability or spikeness of the lightcurve is found to be
correlated with peak luminosity or total isotropic energy of the burst
({\it panel a}).  The time lag of individual peaks seen at different
energy bands is observed to be anticorrelated with luminosity for long
bursts ({\it panel b}).  For short bursts, the lag is small or not
measurable. The $E_{\rm peak}$ is also found to be correlated with
$E_{\rm iso}$ for long bursts including XRFs with short bursts as
clear outliers ({\it panel c}).  The total isotropic energy emission
is correlated with duration ({\it panel d}), with short and long
bursts on approximately the same correlation line, albeit with a wide
spread.  Short bursts detected by \swift\ have lower $E_{\rm iso}$ on
average than long bursts.  There is a group of outliers belonging to
the long burst category which are characterized by being significantly
underluminous. These are GRBs: 980425, 031203, 060614.

Numerous researchers have studied ways to determine the absolute
luminosity of a GRB using correlations such as those illustrated in
Figure \ref{figIIIb}.  These include the lag, variability and $E_{\rm
  peak}$ correlations discussed above.  Other interesting correlations
have included $E_{\rm peak}$ versus $E_{\gamma}$ (emitted energy
corrected for beaming, \citealt{ghirlanda04}) and $E_{\rm peak}$
versus a duration corrected peak luminosity \citep{firmani06}.  The
goal is to derive a method to determine the burst luminosity
independently of a redshift distance determination, thus attempting to
make GRBs standard candles which could be used, in principle, to
determine the cosmic expansion history of the universe to higher
redshift than with supernovae.  Although such efforts are currently
underway \citep[and references therein]{schaefer06}, it is not clear
at present whether any of these correlations are tight enough for
significant progress to be made \citep{bloom03}.

\subsubsection{Soft $\gamma$ repeaters and short bursts.}
It has been noted (\citealt{Hurley:2005:EBF},
\citealt{Palmer:2005:GRF}, \citealt{Nakar:2006:DSH}) that the giant
flare (GF) observed from the putative galactic magnetar source
SGR1806-20 in December 2004 (\citealt{2005Natur_434_1104G}) could have
looked like a typical short GRB had it occurred much farther away,
thus making the tell-tale periodic signal characteristic of the
neutron star rotation in the fading emission undetectable. The two
previously recorded GFs of this type, one each from SGR 0520$-$66 on 5
March 1979 \citep{fenimore96} and SGR 1900$+$14 on 27 August 1998
\citep{hurley99}, would have been detectable by existing instruments
only out to $\sim$ 8 Mpc, and it was therefore not previously thought
that they could be the source of short GRBs. The main spike of the 27
December event would have resembled a short, hard GRB if it had
occurred within $\sim40$ Mpc, a distance scale encompassing the Virgo
cluster \citep{Palmer:2005:GRF}.  However, the paucity of observed GFs
in our own Galaxy has so far precluded observationally based
determinations of either their luminosity function or their rate. The
observed isotropic distribution of short BATSE GRBs on the sky and the
lack of excess events from the direction of the Virgo cluster suggests
that only a small fraction, $\leq$ 5\%, of these events can be SGR GFs
within 40 Mpc \citep{Palmer:2005:GRF}.

Before \swift\ detected short GRBs and their associated afterglow
signatures, searches for nearby galaxies within narrow Inter Planetary
network (IPN) error boxes revealed that only up to $\simeq $ 15\% of
them could be accounted for by SGRs capable of producing GFs
\citep{Nakar:2006:DSH}.  A recent, intriguing candidate is short GRB
070201 which was observed by the IPN to have a location consistent
with the arms of the nearby (0.8 Mpc) M31 galaxy
\citep{2008ApJ_680_545M}.  A LIGO search for gravitational waves
\citep{2008ApJ_681_1419A} at the time of the burst turned up with no
signal, thus excluding a compact merger origin.  If the GRB was really
in M31, it may have been an SGR GF. While the fraction of SGR events
among what are now classified as short GRBs may not be dominant, it
should be detectable and can be tested with future
\swift\ observations.
   It is also worth noting that some short GRBs likely originate in
   the local univserse \citep{2005Natur.438..991T}.

\subsection{Afterglow Observations}

\subsubsection{X-ray observations.}
\begin{figure}
\begin{center}
\includegraphics[width=5.5in]{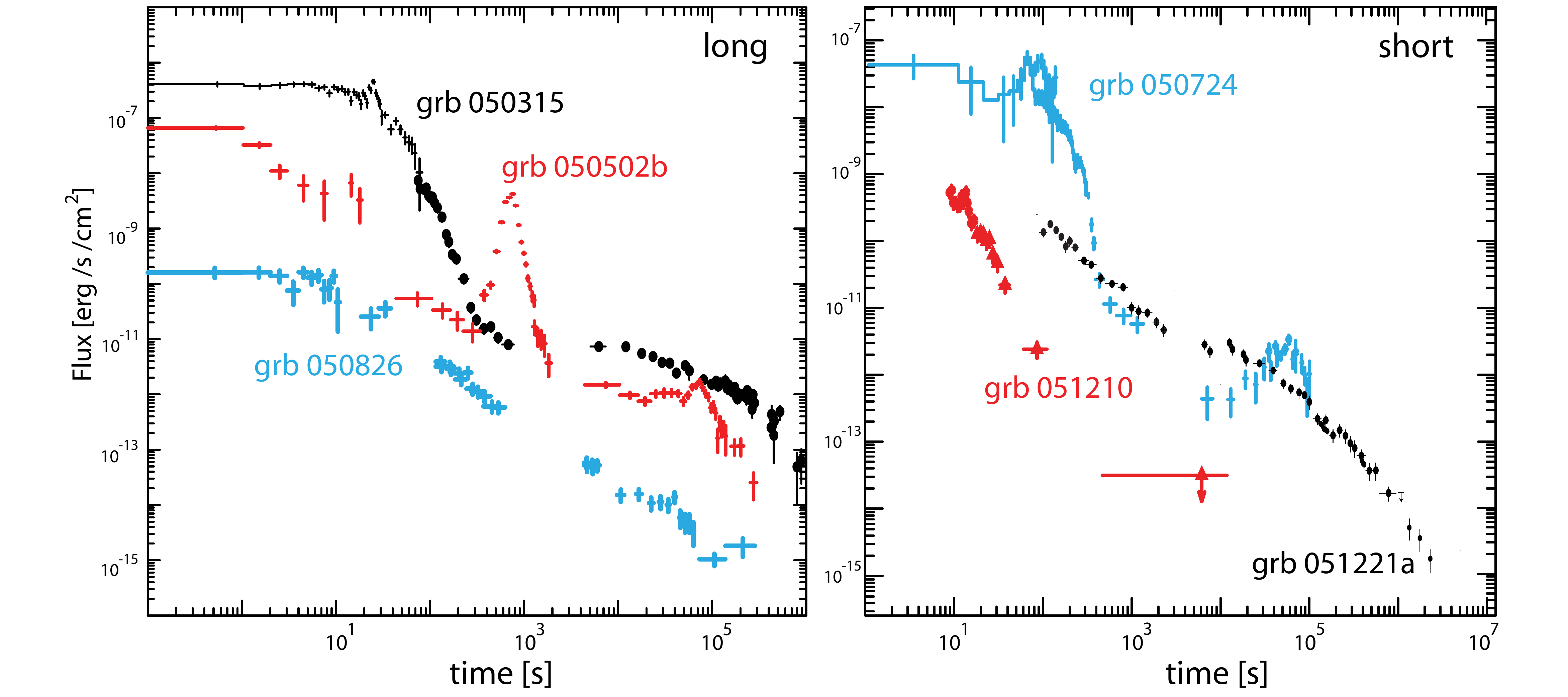}
\end{center}
\caption{\small
  Representative examples of \xray\ afterglows of long
  and short \swift\ events with steep-to-shallow transitions (GRB
  050315, 050724), large \xray\ flares (GRB 050502B, 050724), fastly
  declining (GRB 051210) and gradually declining (GRB 051221a, GRB
  050826; flux scale divided by 100 for clarity) afterglows.  }
\label{figIIIc}
\end{figure}

\swift\ was designed to investigate the GRB afterglows by filling the
temporal gap between observations of the prompt emission and the
afterglow.  The combined power of the BAT and XRT has revealed that
prompt \xray\ emission smoothly transitions into the decaying afterglow
\citep{2005ApJ_635L_133B,obrien06}.  Three representative
\swift\ \xray\ lightcurves are shown in Figure \ref{figIIIc} for both
long and short GRBs. These \xray\ light curves start as early as a
hundred seconds after the GRB trigger, and cover up to five decades in
time. The complex behavior revealed in them sigificantly challenges
traditional afterglow theoy, and calls into question some of the basic
underlying assumptions.

One of most striking results is that many of the early X-ray
afterglows show a canonical behavior, where the light curve broadly
consists of three distinct power law segments \citep{nkg+06}.  A
bright rapid-falling ($t^{-\alpha}$ where $\alpha>3$) afterglow
immediately after the prompt emission \citep{2005Natur.436..985T} is
followed by a steep-to-shallow transition, which is usually
accompanied by a change in the spectrum power-law index.  This is
consistent with an interpretation \citep{nkg+06,2006ApJ_642_354Z} in
which the first break occurs when the slowly decaying emission from
the forward shock becomes dominant over the steeply decaying tail
emission of the prompt $\gamma$-rays as seen from large angles
\citep{2000ApJ_541L_51K}.  Since these two components arise from
physically distinct regions, their spectrum would generally be
different. The shallow phase then transitions to the classical
afterglow phase with no clear evidence for a spectral change.  In some
cases a jet break is seen at late times. The intermediate shallow flux
stage is commonly interpreted as being caused by the continuous energy
injection into the external shock \citep{nkg+06,2006ApJ_642_354Z}
although orientation and complex jet structures have been also
discussed as viable alternatives. 
The energy in the afterglow at these late times is estimated to be
comparable to or smaller than that in the prompt gamma-ray emission,
even when correcting for radiative losses from the afterglow shock at
early times, implying a high efficiency of the prompt emission.  The
presence of the shallow decay phase implies that most of the energy in
the afterglow shock was either injected at late times after the prompt
gamma-ray emission was over or was originally in slow material that
would not have contributed to the prompt gamma-ray emission. This
requires the prompt gamma-ray emission mechanism to be significantly
more efficient than previous estimates. If a significant fraction of
the radiated energy goes to photon energies above the observed range,
the efficiency requirements of the prompt emission become even more
severe.

\begin{table}[!th]
\caption{Typical parameters of the canonical \swift\ \xray\ light
  curve\label{tableIII}} 

\vspace*{0.1in}

\begin{tabular}{lccc}\hline
~     & ~           & ~               & Approximate \\
Phase & Start T (s) & Decay index$^a$ & frequency \\ \hline
Steep decline & $10{^1}-10{^2}$ & $>$3 & 50\% \\
Shallow slope & $10{^2}-10{^3}$ & 0.5 & 60\% \\
Classical afterglow & $10{^3}-10{^4}$ & 1.3 & 80\% \\
Jet break late phase & $10{^5}-10{^6}$ & 2.3 & \phantom{$^b$}20\%$^b$ \\
X-ray flares & $10{^2}-10{^4}$ & & 50\% \\ \hline
\end{tabular}

\begin{tabular}{p{4.4in}}
\small%
$^a$Decay index $\alpha$ defined by $F = F_o t^{-\alpha}$. \\
\small%
$^b$Of the 80\% with no observed jet break, about half had afterglow
observations terminate before expected time of jet break. \\
\end{tabular}
\end{table}

\begin{figure}[p]
\begin{center}
\includegraphics[scale=0.6]{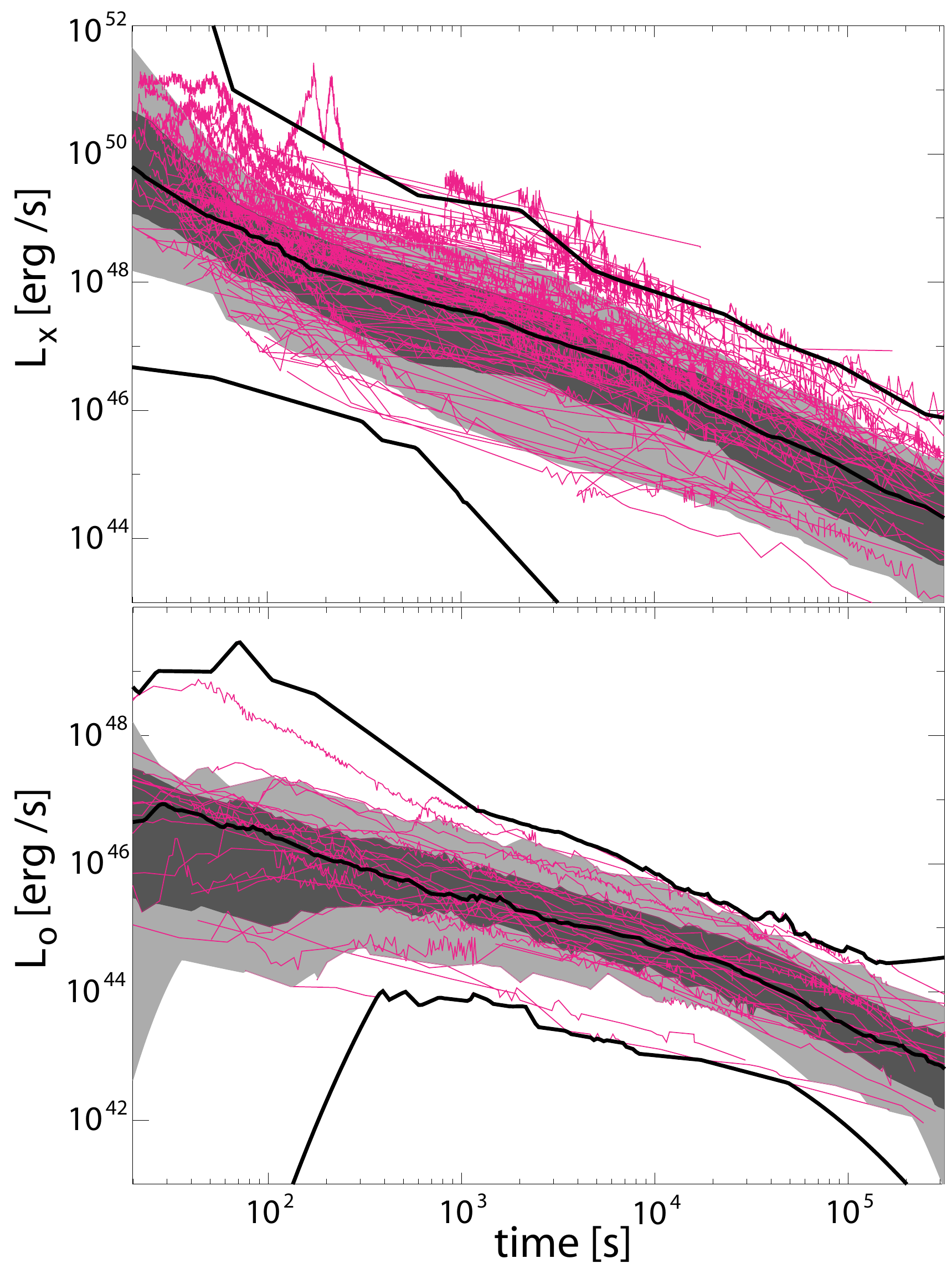}
\end{center}
\caption{\small
  \xray\ and optical lightcurves of GRB afterglows in
  the \swift\ era.
  ({\it a}) \xray\ lightcurves of
  \swift\ burst afterglows.  Data for long-duration bursts with known
  redshifts, from GRB\,050126 to GRB\,070724A, have been gathered from
  the \swift\ XRT lightcurve and spectral data depository at the UK
  \swift\ Science Data Centre (\citealt{ebp+07};
  \citealt{ebp+09}). They are corrected uniformly to unabsorbed
  luminosity over $1.0-30.0$ keV in the burst rest-frame, using the
  time-average afterglow spectrum, and plotted as a function of
  rest-frame time (pink lines).  Separately, afterglow lightcurve fits
  \citep{rlb+09}, which exclude flaring intervals, are used to
  construct minimum and maximum envelopes (black lines) and confidence
  intervals (gray bands) on the \xray\ luminosities of the bursts as a
  function of rest-frame time: light gray regions delimit bands of
  10\% to 90\% confidence, dark gray regions delimit bands of 25\% to
  75\% confidence, and the median burst luminosity at any given time
  is shown (middle black line).
  ({\it b}) Optical lightcurves of \swift\ burst afterglows.  Data for
  long-duration bursts with known redshifts and at least ``bronze''
  quality published optical data \citep{kkz+07}, from GRB\,050408 to
  GRB\,070612A, are corrected uniformly to rest-frame $U$-band
  luminosity using the inferred $R$-band ($z=1$) light-curves from
  \citet{kkz+07}, and plotted as a function of rest-frame time (pink
  lines).  Interpolated and ``best fit'' extrapolated lightcurves are
  used to generate minimum and maximum envelopes and median luminosity
  estimates (black lines) and confidence intervals (gray regions), as
  in \textit{a}.}
\label{figIIId}
\end{figure}

The average times, slopes, and frequencies characterizing these three
distinct \xray\ afterglow components are listed in
Table~\ref{tableIII}.  Most \swift\ \xray\ light curves are broadly
consistent with this basic temporal description, although in most
cases we do not see all three power law segments, either because not
all are present or because of limited temporal coverage.  The large
variety of behaviors exhibited by afterglows at different times in
their evolution can be seen in Figure~\ref{figIIId}, which shows the
temporal history for each individual afterglow as well as the
evolution of the cumulative \xray\ afterglow luminosity for a large
sample of \swift\ events with known redshift.  While broadly
compatible with relativistic fireball models
\citep{nkg+06,2006ApJ_642_354Z} the complex afterglow behavior that
has been revealed poses new challenges of interpretation. The reader
is refered to \citet{2008arXiv0811_1657G} for a more detailed account
of the major strengths and weaknesses of the standard afterglow model,
as well as some of the challenges that it faces in explaining recent
data.

\swift\ has also discovered flaring behavior appearing well after the
prompt phase in $\sim$50\% of \xray\ afterglows
\citep{2007ApJ_671_1903C,2007ApJ_671_1921F}.  An illustration of
bursts with bright flares is shown in Figure~\ref{figIIIc}.  In some
extreme cases, the late time flares have integrated energy similar to
or exceeding the initial burst of $\gamma$-rays \citep{brf+05}.  The
rapid rise and decay, multiple flares in the same burst, and cases of
fluence comparable to the prompt emission suggest that these flares
are due to the same mechanism responsible for the prompt emission,
which is usually attributed to the activity of the central powerhouse.
When \xray\ flares are observed by XRT, it is typically the case that
no flaring is seen in the optical band by the UVOT.  A notable example
is GRB 060418 \citep{2007AA_469L_13M}, whose optical-infrared
afterglow spectra is not consistent with a simple power-law
extrapolation to soft \xray\ energies and clearly requires two distinct
spectral components. Not surprisingly, the broadband energy spectra of
GRBs are complex and have spectra of wildly disparate shapes even for
members belonging to a single {\it class}.
		
Prior to \swift\ there were several reports of emission and absorption
line observations in the \xray\ spectra of GRB afterglows.  These
included BeppoSAX observations of GRB 970508 \citep{beppo} and GRB
000214 \citep{antonelli00}, \asca\ observations of GRB 970828
\citep{yoshida99}, \chandra\ observations of GRB 991216 \citep{piro00}
and \xmm\ observations of GRB 011211 \citep{reeves02}.  None of the
detections were of high statistical significance, but, combined, were
of some credibility.  \swift\ has not found any significant line
features in comprehensive observations of the \xray\ afterglow of more
than 200 GRBs \citep{romano08}.  \cite*{2008ApJ...679..587H} also made
a very detailed \swift\ \xray\ line search, with no positive result.
These strong negative findings call into question the significance of
previous results.

A final, but widely discussed, observational development is the
discovery in $\sim$25\% of short bursts detected by \swift/BAT of an
extended emission (EE) component lasting for $\sim$100 seconds
\citep{norris08}.  This component was       clearly detected in
\hete\ burst GRB 050709 \citep{villasenor05} and \swift\ burst GRB
050724 \citep{barthelmy05}.  Archival searches have also found BATSE
bursts with EE (\citealt{norris_bonnell06}, \citealt{2001AA_379L_39L},
\citealt{2002ApJ_567_1028C}).  The EE is typically softer than the
main peak and has an intensity range from 10$^{-3}$ to 10$^{-1}$ times
that of the initial short pulse complex.  It is possible that many of
the 75\% of bursts without currently detected EE have this component
at flux levels below detectability, although there are bursts with
upper limits on the intensity of $<10^{-4}$ times that of the short
pulse complex.
		
\subsubsection{Optical observations.}
With increasing frequency during the \swift\ era, optical observations
of GRBs have been commencing almost immediately after -- and in one
noteworthy case (GRB 080319B; \citealt{rks+08}), prior to -- the GRB
trigger itself.  These early optical observations have substantially
enriched our appreciation for the complexity of the physical processes
active during the prompt emission phase of GRBs, and during the
afterglow that follows. The large diversity of optical afterglow
lightcurves can be seen in Figure~\ref{figIIId}. 

\paragraph{Prompt emission.}
Observations prior to the \swift\ era demonstrated already that bright
optical flashes such as that seen from GRB\,990123 \citep{abb+99},
were relatively rare \citep{kab+01}. The success of the
\swift\ mission has brought a vast increase in the rate of bursts
accessible to rapid optical follow-up, and a corresponding increase in
the number of events detected from early times, $t < 100$\,s.  The
\swift\ UVOT itself routinely detects optical afterglows following the
intial prompt slew of the satellite, with a 40\% detection rate for
such bursts \citep{rko+08} that is only slightly lower than the 60\%
detection rate combining all observatories.

The fastest routine responses to \swift\ alerts are realized by
robotic ground-based telescopes.  The first discovery yielded by these
observatories in the \swift\ era was of the $\gamma$-ray correlated
component of the prompt optical emission (\citealt{vww+05},
\citealt{bbs+05}, \citealt{vww+06}).  This component is not observed
in every burst, but the mere fact of its correlation is sufficient to
establish a common origin with the prompt $\gamma$-ray emission (e.g.,
internal shocks).  When observed, the ratio of the correlated
$\gamma$-ray to optical flux densities has been found to be roughly
$10^5$ to one.

In contrast to bursts with $\gamma$-ray correlated emission, the
common burst is now revealed to either exhibit a single power-law
decay from early times (\citealt{ryk+05}, \citealt{qry+06},
\citealt{yar+06}), or to exhibit a flat or rising (\citealt{rsp+04},
\citealt{rmy+06}) or rebrightening \citep{sdp+07} optical light curve
before it enters the standard power-law afterglow decay.  The initial
brightness of the typical counterpart is $V\sim 14$ to 17~mag\footnote{The
  $\sim60$\% of those counterparts not detected are presumably fainter than this.}
\citep{rko+08}, which has made observations challenging for smaller
($<$1\,m) robotic facilities, and has limited the extent of the
lightcurves collected by the UVOT.
 
A few new observations of bright flaring optical emission have been
collected, which are usually interpreted as emission from the reverse
shock region (\citealt{sp99a}, \citealt{mr99}).  The early optical/NIR
emission from GRB\,041219 would have rivalled that seen from
GRB\,990123 if not for the large Galactic extinction along the line of
sight \citep{vww+05}.  Of the three distinct peaks observed by
PAIRITEL, the second may represent the onset of the afterglow (reverse
shock) contribution \citep{bbs+05} -- if so, a relatively small
Lorentz factor, $\Gamma \sim 70$, is derived by associating the
flaring peak time with deceleration time of the relativistic blast
wave.  Observations of GRB\,050525A with UVOT \citep{bbb+06} and
GRB\,060111B with TAROT (in a unique time-resolved tracking mode;
\citealt{kgs+06}) show the ``flattening'' light-curve familiar from
GRB\,021211 \citep{zkm03}.  Intriguingly, the ``high redshift''
GRB\,050904 (\citealt{2005A&A...443L...1T}, \citealt{cmc+06}) at
$z=6.29$ \citep{kka+06} also had prompt optical emission observed
(\citealt{bad+06}, \citealt{hnr+06}), with a brightness, single-pulse
structure, and fast-fading behavior reminiscent of GRB\,990123 and
thus, potentially also interpreted as reverse shock emission.

\swift\ detection of the ``naked eye burst'' GRB\,080319B, which
peaked at visual magnitude $V = 5.3$\, has now delivered the richest
dataset, by far, addressing the prompt optical emission and its
evolution into a standard fading afterglow (\citealt{rks+08},
\citealt{2009ApJ_691_723B}, \citealt{wvp+08}).  This is only partially
due to the extreme brightness of the event; the fact that it occurred
at an equatorial location, in the night sky above the Western
hemisphere, is probably even more important.  The fact that it
occurred within just one hour and ten degrees of the preceding
GRB\,080319A means that it is also the only event with strong
contraints on optical precursor emission from pointed telescopes
\citep{rks+08}.  The GRB\,080319B dataset is rich enough that its
ramifications are still being grappled with.  In an overall sense the
prompt optical emission correlates well with the $\gamma$-ray
lightcurve; in detail, though, the individual pulses observed in both
bands do not track precisely, probably suggesting the presence of at
least two distinct cooling processes at play within the dissipation
region \citep{rks+08}.
 GRB\,080319B also allowed astronomers to rule out inverse comptonization
  as a relevant mechanism (\citealt{2009MNRAS.393.1107P}, 
                           \citealt{2009ApJ...692L..92Z}).\\

\paragraph{Afterglow emission.}
The observed properties of the optical afterglow are largely familiar
from observations prior to the \swift\ era (\citealt{vkw00},
\citealt{meszaros02}) and will not be reviewed in detail here.  An
intriguing feature of later-time afterglows, revealed by the rich
multiband lightcurves available in the \swift\ era, has been the
occasional presence of chromatic lightcurve breaks, where the X-rays
show a clear break (steepening of the flux decay rate) while the
optical does not. The break in the \xray\ light curve is usually
identified with the end of the shallow decay phase. The optical
lightcurve follows a single power law decay, usually with a temporal
decay index intermediate between those in the X-rays before and after
the break. Accommodating such chromatic breaks in a model where the
\xray\ and optical emissions arise from the same emitting region
\citep{pmb+06} requires not only a temporal evolution of the
underlying microphysical conditions within the emitting region but
also fine tuning their photon arrival times in such a way that a break
in the \xray\ will be produced but not at optical
wavelengths. Alternatively, the \xray\ and optical photons may arise
from physically distinct regions, which would naturally account for
their seemingly decoupled behavior. Observations of the naked-eye
GRB\,080319B have sharpened this debate. This is because the optical
and \xray\ decays in this event exhibit discrepant behavior over two
orders of magnitude in time, from 100\,s to a few times $10^4$\,s
post-burst (\citealt{rks+08}, \citealt{kp08}).


Recent results showed that during the initial 500s of observations
15\% of UVOT lightcurves are seen to rise - with an average peak time
of 400s, 58\% decay from the onset of observations and the remainder
are consistent with being flat, but could be rising or decaying
\citep{2009MNRAS.tmp..376O}. This leads to a wide range of temporal
indices measured before 500s, $-1.17<\alpha<0.21$.
       Such behavior is also observed by ground
based telescopes (\citealt{2005ApJ...631.1032R}, 
                  \citealt{2006ApJ...640..402Q},
                  \citealt{2006ApJ...636..959Y},
                  \citealt{2006ApJ...638L...5R})
          and was seen in pre-\swift\  observations 
                  \citep{2004ApJ...601.1013R}.
No color evolution was observed during the rising phase of the UVOT
lightcurves. One likely scenario is that the rise is caused by the jet plowing into
the external medium - the start of the forward shock. If this scenario
is correct, then it is possible to determine the Lorentz factor of the jet at
the time of the peak, giving $\Gamma\sim180$ and a lower limit $\Gamma>230$
for those without observed peaks \citep{2009MNRAS.tmp..376O}.

After a few hundred seconds the afterglow decays as a single power-law
with a temporal index, measured after 500s, of $-1.20<\alpha
<-0.52$. This is consistent with pre-\swift\ observations and the
range is similar to the shallow decay of the XRT canonical
model. However, 20\% of UVOT lightcurves are seen with a broken
power-law decay after 500s. For these afterglows, the first decay has
a range between $-0.74<\alpha<-0.46$, which is most consistent with
the shallow decay of the \xray\ canonical lightcurve and the second
decay has a temporal range of $-1.72<\alpha<-1.34$, which is most
consitent with the classic afterglow phase.

At late times, $t\simgt 1$\,day, it was common in the pre-\swift\ era
to observe a steepening of the optical decay to power-law indices
$\alpha_{\rm o}>2$, and to associate this epoch with the ``jet break''
transition of the underlying, relativistically expanding jet.  As jet
breaks are intrinsically achromatic, they are expected to manifest in
the \xray\ light curves from \swift\ as well as in ground-based
optical observations.  In fact, over the first three years of
\swift\ operations only a handful of convincing jet-break candidates
were identified (\citealt{bbb+06}, \citealt{sdp+07}, \citealt{dhm+07},
\citealt{willingale07}, \citealt{kocevski08}), leading to concerns
about the viability of this picture.
   Some bursts with very long coverage
   have been convincingly shown to 
  possess  no significant breaks, e.g., 
  GRB\,060729 \citep{2007ApJ...662..443G}.

Since then, however, deep optical imaging observations have revealed
evidence for jet breaks in several additional \swift\ afterglows, a
significant fraction of those monitored to late times \citep{dgp+08}.
Systematic analysis of the \swift\ XRT data has revealed strong
evidence for a jet-break signature in 12\% of the \xray\ lightcurves
\citep{rlb+09}.  A consensus has developed that jet breaks for typical
\swift\ bursts may be occurring at late times and faint flux levels
that are beyond the limit of the standard ground- and space-based
campaigns.  Separately, it has been suggested that a distinct
additional spectral component could hide the jet-break signature in
the \xray\ band. For example, in the case of the bright GRB\,070125
\citep{ccf+08}, inverse Compton scattering of the synchrotron optical
photons has been put forward as an explanation for the missing
jet-break feature.

\subsubsection{Radio observations.}
Radio afterglow observations are unique in having led to both indirect
and direct demonstrations of relativistic expansion, via scintillation
(e.g. \citealt{ccf+08}) and VLBI \citep{tmp+05} observations.  They
provide access to the properties of bright afterglows on the smallest
angular scales. The limited number of sensitive high-resolution radio
facilities, and the sensitivity limits of those facilities, have
prevented a proportional exploitation of the greatly-increased burst
rate from \swift, primarily because the compensating feature of this
increased burst rate has been a greater median redshift and a lower
characteristic afterglow flux.

Radio afterglow observations nonetheless continue to play a vital role
in accurately estimating blastwave kinetic energies
(\citealt{2004MNRAS_353L_35O}, \citealt{2005ApJ_618_413G},
\citealt{2007ApJ_654_385K}), with radio detections contributing
crucially to the demonstration of the extremely large ($E\sim
10^{52}\,$erg) kinetic energy associated with the high-redshift
GRB\,050904 \citep{fck+06,gfm07}, and to constraining the relativistic
energy associated with the nearby GRB\,060218/SN\,2006aj
\citep{skn+06}. Radio data can also provide a crucial ``third check''
on claims of jet-break detections, as with the broadband afterglow
models applied to GRB\,050820 \citep{ckh+06} and GRB\,070125
\citep{ccf+08}.

\begin{figure}
\begin{center}
\includegraphics[scale=0.55]{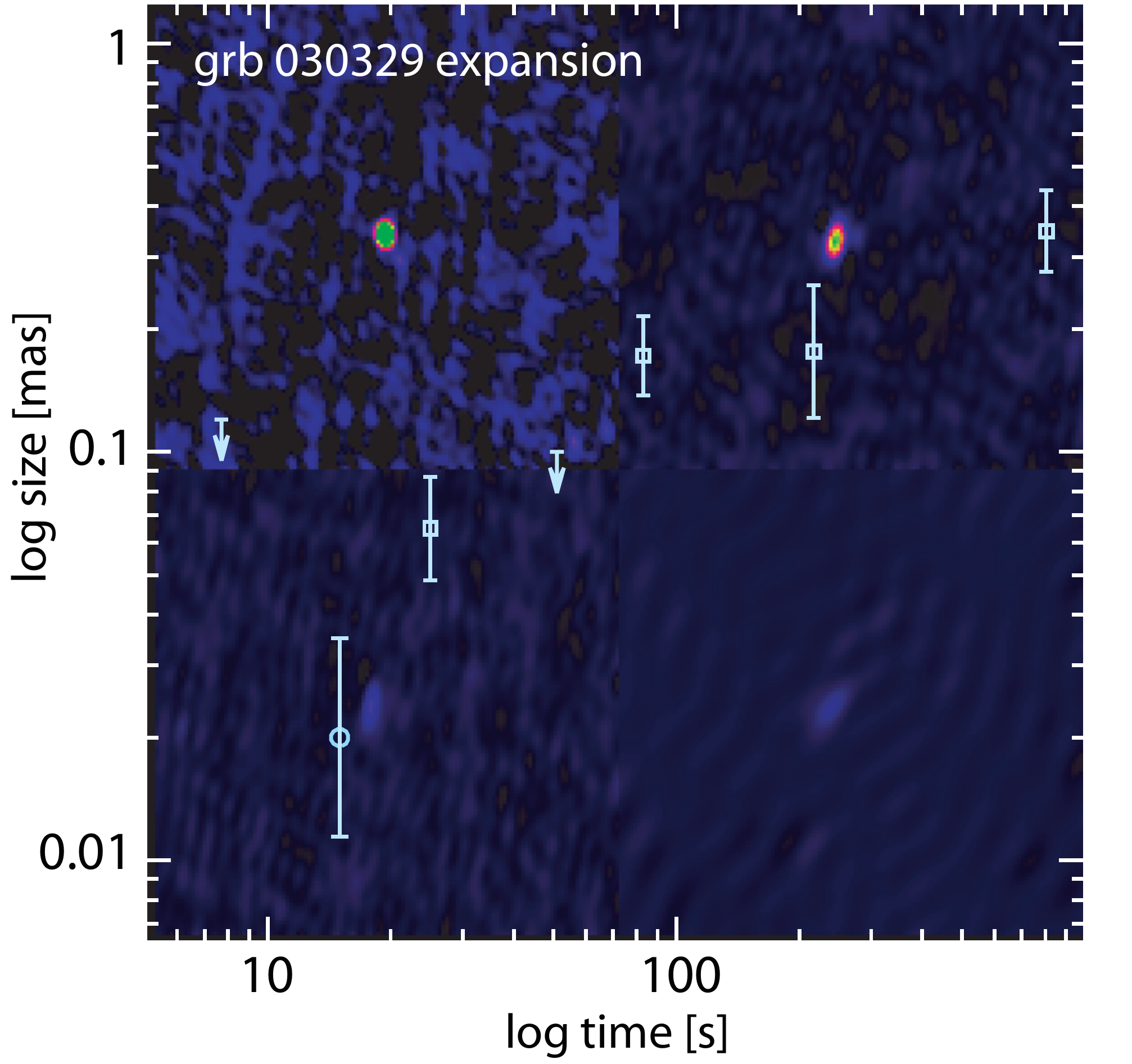}
\end{center}
\caption{\small
  The growth of GRB\,030329 with time as measured using VLBI (by
  \citealt{2007ApJ_664_411P} and references therein).  In the
  background are the images from (a) April 2003 (15 GHz), (b) June
  2003 (8 GHz), (c) November 2003 (8 GHz), and (d) June 2005 (5 GHz)
  with the same intensity scale.  The resolution for the four images
  is not constant in time, but is accounted for in the analysis of the
  source size.  }
\label{exp030329}
\end{figure}

For relatively nearby GRBs which may be associated with a supernova,
radio observations have proven invaluable in providing evidence of the
physical expansion of GRB/SNe ejecta through direct imaging. Figure
\ref{exp030329} shows the expansion from $\sim0.02$ mas to $\sim0.35$
mas of the radio image of GRB 030329 over a time span from $\sim15$s
after the GRB trigger to $\sim10^3$s \citep{2007ApJ_664_411P}.

\subsection{Interpreting Prompt and Afterglow Emission}

The isotropic-equivalent luminosity of GRB \xray\ afterglows scaled to
$t \sim 11$ hours after the burst in the source frame, can be used as
an approximate estimator for the energy in the afterglow shock for the
following reasons (\citealt{2001ApJ_547_922F},
\citealt{2001ApJ_560L_167P}, \citealt{2008ApJ_689_1161G}).  First, at
11 hr the \xray\ band is typically above the two characteristic
synchrotron frequencies, so that the flux has very weak dependence on
microphysical parameters and no dependence on the external density,
both of which are associated with relatively large
uncertainties. Second, at 11 hr the Lorentz factor of the afterglow
shock is sufficiently small ($\Gamma \sim 10$) so that a large
fraction of the jet is visible (out to an angle of $\sim \Gamma^{-1}
\sim 0.1$ rad around the line of sight) and local inhomogeneities on
small angular scales are averaged out. Finally, the fact that the
ratio of $L_X (11\; {\rm hr})$ and $E_{\rm iso}$ is fairly constant
for most GRBs, suggests that both can serve as a reasonable measure of
the isotropic-equivalent energy content of the ejected outflow.

\begin{figure}
\begin{center}
\includegraphics[scale=0.55]{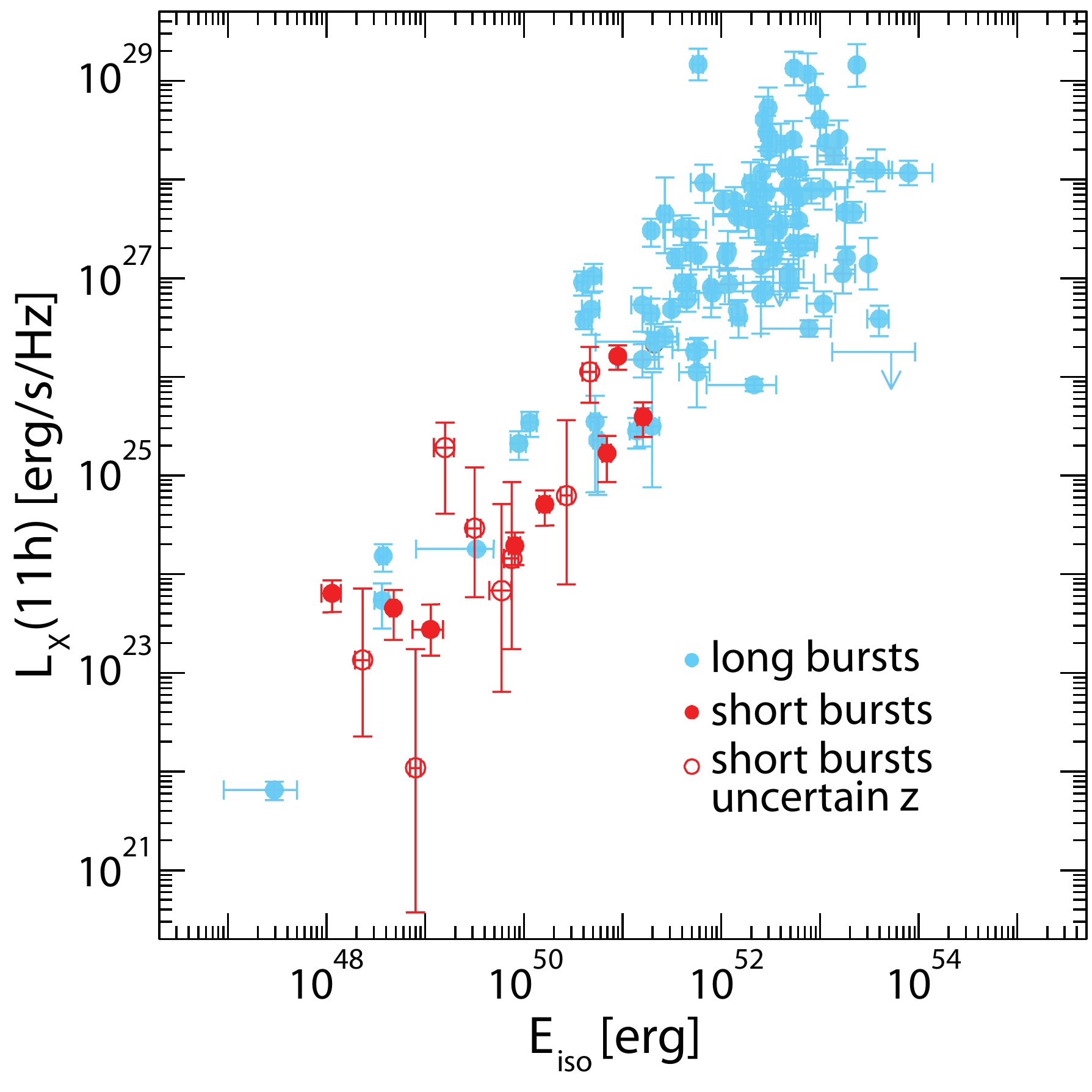}
\end{center}
\caption{\small
  Isotropic-equivalent luminosity of GRB X-ray
  afterglows scaled to $t = 11$ hr (5 keV source frame) after the
  burst trigger as a function of their isotropic $\gamma$-ray energy
  release (adapted from \citealt{2008arXiv0806_3607N}).  }
\label{figIIIe}
\end{figure}

Figure \ref{figIIIe} shows $L_X (11\;{\rm hr})$ at 5 keV rest-frame
energy as a function of their isotropic $\gamma$-ray energy release
for a large sample of GRBs. A linear relation, $L_X \propto (11\;{\rm
  hr}) \propto E_{\rm \gamma,iso}$, seems to be broadly consistent
with the data, probably suggesting a roughly universal efficiency for
converting kinetic energy into $\gamma-$rays in the prompt emission
for both long and short GRBs (\citealt{2005ApJ...630L.165L}, \citealt{nkg+06},
\citealt{2007ApJ_654_878B}, \citealt{2007ApJ_654_385K},
\citealt{2008arXiv0806_3607N}).  This ``universal'' efficiency is also
likely to be high (i.e., the remaining kinetic energy is comparable
to, or even smaller than, the energy dissipated and radiated in the
prompt emission). If this is the case, the well-known efficiency
problem for long GRBs also persists for short events.

\subsection{The Radiated Energy Inventory}

\begin{figure}
\flushleft\hspace*{-0.8in}\includegraphics[width=6.5in]{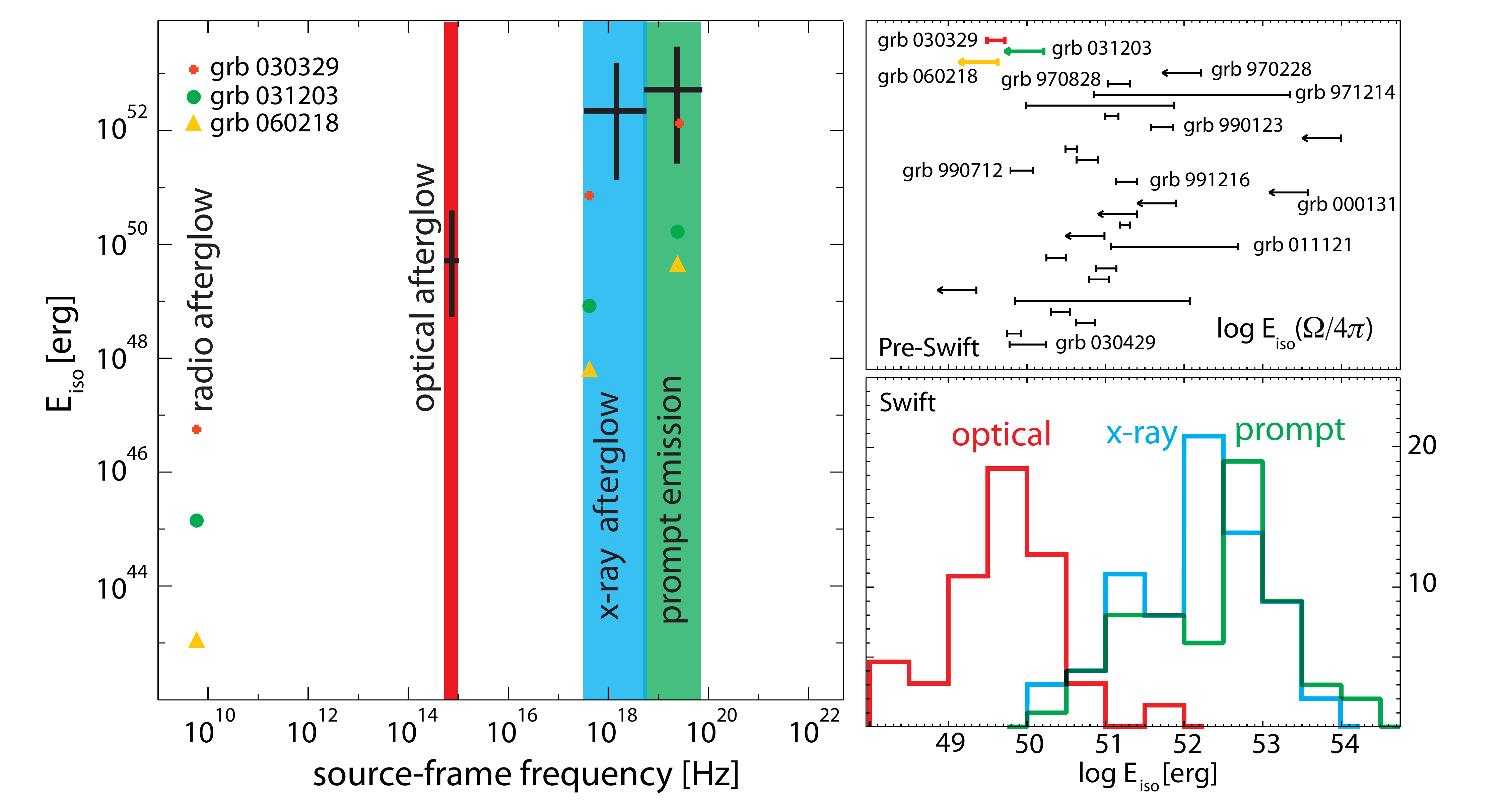}
\caption{\small\setlength{\baselineskip}{0.6\baselineskip}%
   The radiated-energy inventory of \swift\ GRBs.  {\it
    Left Panel:} Summary of the isotropic-equivalent total emitted
  energy of the prompt and afterglow emission for \swift\ GRBs in the
  source frame (adapted from \citealt{2007ApJ_654_385K}).  {\it Right
    Panel Bottom}: Distributions of cumulative isotropic-equivalent
  total emitted energy for \swift\ GRBs.  {\it Right Panel Top}:
  Comparisons of collimation-corrected total emitted $\gamma$-ray
  energy, $E_{\rm iso}(\Omega/4\pi)$, of pre-\swift\ GRBs, where
  $E_{\rm iso}$ is used as an upper limit for GRBs with no jet angle
  constraints.  }
\label{figIIIf}
\end{figure}

With the advent of \swift, the observational inventory for GRBs has
become rich enough to allow estimates of their radiated energy
content. A compilation of the radiated energy in both the prompt and
afterglow phases is presented in Figure \ref{figIIIf}. To investigate
the energy dissipation behavior in the \xray\ and optical afterglow, we
fitted a natural cubic spline function to the afterglow histories
(shown in Figure \ref{figIIId}) for each individual afterglow and
estimated the cumulative emitted energy. The start and end times of
the integration were the first and the last points of the actual
observations. The result is a global portrait of the effects of the
physical processes responsible for GRB evolution, operating on scales
ranging from AU to parsec lengths.  The compilation also offers a way
to assess how well we understand the physics of GRBs, by the degree of
consistency among related entries.

Figure \ref{figIIIf} shows that the isotropic \xray\ emission, $E_{\rm
  X,iso}$, for most \swift\ GRBs spans the range of $\sim 10^{50}$ to
$10^{54}$ erg, which is comparable to that emitted during the prompt
$\gamma$-ray phase. Not surprisingly, events that have large
isotropic-equivalent energy in $\gamma$-rays have larger $E_{\rm
  X,iso}$, indicating a reasonably narrow spread in the efficiency of
converting the afterglow kinetic energy into radiation. As can be seen
also in Figure \ref{figIIIf}, the isotropic equivalent energy that is
radiated at optical wavelengths is $\sim 2$ orders of magnitude
smaller than that in X-rays and $\gamma$-rays. This is predominantly
due to the fact that $\nu F_\nu$ typically peaks closer to the X-rays
than to the optical, and it is very flat above its peak while it falls
much faster toward lower energies. Finally, since these are isotropic
equivalent energies, most of the contribution to the radiated
afterglow energy, especially at radio wavelengths, is from
significantly later times than for $E_{\rm X,iso}$, and the
collimation of the outflow together with relativistic beaming effects
could result in much larger $E_{\rm X,iso}$ than $E_{\rm R,iso}$.

Of the three nearby SN-GRB events plotted in Figure \ref{figIIIf},
only GRB 030329 falls within the $\sim10^{50}$ erg range, while the
other two events fall between $\sim 10^{48}$ and $\sim 10^{49}$ erg
\citep{2007ApJ_654_385K}.  We note that there exists a selection
effect based on the observed photon flux: an event is more likely to
be detected when it is closer to us than farther, for a given
intrinsic luminosity.

One of the liveliest debated issues associated with GRBs concerns the
total energy released during the explosion: are GRBs standard candles?
The GRB community has vacillated between initial claims that the GRB
intrinsic luminosity distribution was very narrow
\citep{1994ApJ_429_319H} to discounting all standard-candle claims, to
accepting a standard total GRB energy of $\sim 10^{51}$ ergs
\citep{fks+01}, and to diversifying GRBs into “normal” and
“sub-energetic” classes (\citealt{2008Natur_453_469S},
\citealt{2005ApJ_631_435R}).  The important new development is that we
now have significant observational support for the existence of a
sub-energetic population based on the different amounts of
relativistic energy released during the initial explosion. A network
of theoretical tests lends credence to this idea
(\citealt{2004ApJ_605L_97W}, \citealt{2007ApJ_654_385K},
\citealt{2004ApJ_609L_9G}, \citealt{2006MNRAS_366L_13G},
\citealt{2006Natur_442_1011P}).  The existence of a wide range of
intrinsic energies could further challenge the use of GRBs as standard
candles.


\section{ENVIRONMENTS AND HOST GALAXIES}
\label{sec:hosts}

Much of what we know about GRBs has been derived not from observations
of the prompt burst radiation itself, but from studies of their
afterglows -- as they illuminate the circumburst surroundings -- and
their host galaxies.  In the sections that follow we discuss the
primary insights that have been derived in this manner from the study
of GRB environments.


\subsection{Cosmological Setting}
\label{sub:hosts:intro}

In Figure~\ref{figIIIa} we present the redshift distribution of all
\swift-detected gamma-ray bursts. \swift\ and other current missions
observe GRBs to cosmological distances quite readily; indeed, the
three highest-fluence, known-redshift bursts observed by \swift\ have
been at $z=0.61$ (GRB\,050525A), $z=2.82$ (GRB\,050603), and $z=1.26$
(GRB\,061007) -- already spanning 40\% of cosmic history.
Historically, the majority of redshifts have been collected via host
galaxy spectroscopy; in the \swift\ era, however, this pattern has
been reversed -- except for the short bursts -- with the great majority
of redshifts now being derived from afterglow spectra.   

In addition to theoretical arguments that posit different physical
origins for short and long bursts (e.g., \citealt{1996ApJ_471_915K}),
the absence of short burst-associated supernovae to deep limits
(\citealt{hjorth05b}, \citealt{ffp+05}, \citealt{2006AIPC_836_473B},
\citealt{castro-tirado05}), 100 times fainter than SN\,1998bw in the
best cases, argues for a distinct origin of the short and long bursts.
In agreement with this picture, the redshift distributions of the two
populations are not consistent.

For the long bursts, which are associated with active star formation
and, in particular, the deaths of massive stars, it is interesting to
explore whether their distribution in redshift is consistent with
other measures of cosmic star formation.  The greatly-increased number
of redshifts available for \swift\ bursts 
have motivated several such comparisons, and for the
first time, estimates of cosmic star formation at high redshift $z>4$
using the \swift\ redshift sample \citep{cbc07,ykb+08}.

The star-formation studies over $1<z<4$ confirm, in a broad sense,
that the GRB redshift distribution remains consistent with independent
measures of star formation \citep{jlf+06a}.  However, there are signs
of differential evolution of the GRB rate, in the sense that the GRB
rate increases more rapidly with increasing redshift than expected
based on star-formation measures alone (\citealt{ld07},
\citealt{gp07}, \citealt{kyb+08}).  This evidence, currently present
at roughly 95\%-confidence, may strengthen significantly in coming
years.  If so, this would provide a       sign of bias towards
low-mass and low-metallicity host galaxies -- and potentially,
low-metallicity progenitors -- for the long-duration bursts (Section
\ref{sub:hosts:long}). 
  The theoretical curves accompanying the GRB redshift distribution
(Figure~\ref{figIIIa})
 which show the  evolution of a comoving volume element in the universe,
 and the volume convolved with star formation rate,
    appear to indicate  that the observed distribution
    is wider than expected. 

The short-burst redshift distribution, so far drawn exclusively from
host-galaxy observations, has also been compared to star-formation
metrics.  In this case, however, the intent has been to explore
``time-delayed'' progenitor models that correlate with star formation
through a parametrized (log-normal or power-law) delay function
(\citealt{ngf06}, \citealt{gp06}, \citealt{scc+08}).  Consistent with
the relatively large fraction of events at $z\simlt 0.5$ compared to
long bursts, and with host galaxy demographics
(\citealt{zr07}, \citealt{sb07}), these studies have concluded that a
long-lived ($\tau\simgt 1$\,Gyr) progenitor is required for these
models to be consistent with \swift-era redshift measurements and the
distribution of short burst fluences from BATSE.  This in turn has led
to relatively high estimates of the volumetric local short burst rate,
at least an order of magnitude greater than the local rate of long
bursts \citep{Guetta:2005:CSG}, and a correspondingly optimistic set
of predictions for Enhanced LIGO, VIRGO, and other ground-based
gravity-wave detectors.


\subsection{Host Galaxies of Long Bursts}
\label{sub:hosts:long}

\begin{figure}
\begin{center}
\includegraphics[width=5.5in]{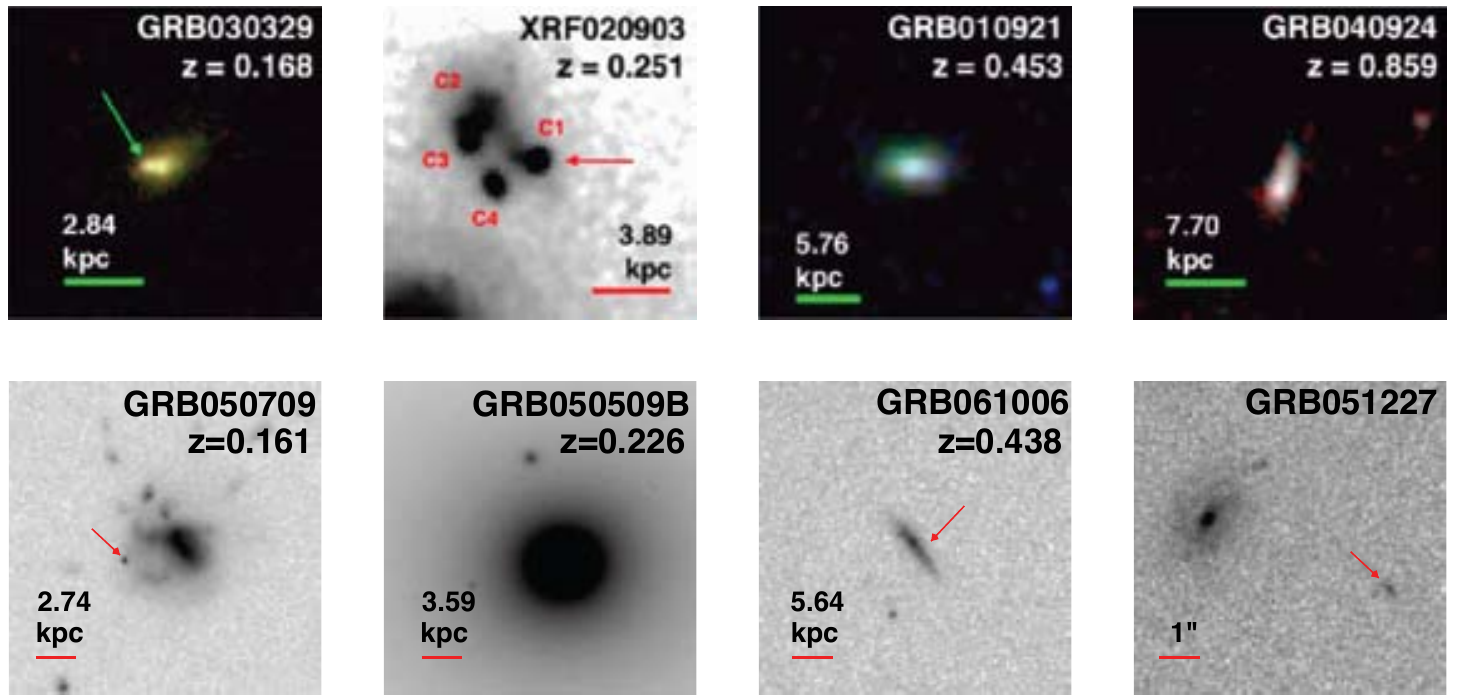}
\end{center}
\caption{\small
   A selection of the host galaxies of long-duration
  (top row) and short-duration (bottom row) gamma-ray bursts, as
  imaged by the \textit{Hubble Space Telescope}.  An attempt has been
  made to choose pairs of long and short burst host galaxies with
  comparable redshifts; lower-redshift hosts are emphasized as these
  reveal their structure more readily in short exposures.  Images are
  oriented with North up and East to the left, and the physical length
  scale for a one-arcsecond angular distance is indicated in each
  panel (except for GRB\,051227); arrows point to the location of the
  burst where this is known to $\sim$pixel precision.  Individual
  burst notes: GRB\,030329 was the first classical long GRB to be
  associated with a well-observed spectroscopic supernova
  (\citealt{2003ApJ_591L_17S}, \citealt{Hjorth:2003:VES}); XRF\,020903
  was the first X-ray Flash event to yield a redshift measurement
  \cite{Soderberg:2004:RDX}; GRB\,050709 was the first short burst
  with optical afterglow -- indicated by the arrow -- detected
  \cite{hwf+05,ffp+05}; GRB\,050509B was the first short burst with
  detected afterglow (\citealt{2005Natur_437_851G}; 
                      \citealt{2006ApJ_638_354B})
    GRB\,051227 has a
  faint candidate host, of unknown redshift probably greater than 1,  visible at
  the optical afterglow location; the spiral galaxy to the east has
  redshift $z=0.714$ \cite{2005GCN_4409_1F}.  Long-burst host images
  from \cite{wbp07}; short-burst host images from \cite{ffp+05} and
  this work.  }
\label{figIVa}
\end{figure}

Surveys of GRB host galaxies in the pre-\swift\ era (\citealt{ldm+03},
\citealt{fjm+03}) necessarily focused on the host galaxies of
long-duration bursts.  These surveys established a standard picture
for the GRB hosts as sub-\Lstar\ galaxies (median $L\sim 0.1$\Lstar)
with exponential-disk light profiles \citep{cvf+05,wbp07} and high
specific star-formation rates (SSFR $\sim$ 1~Gyr$^{-1}$)
\citep{chg04}.  A selection of GRB host galaxies, as imaged by \hst,
are shown in Figure~\ref{figIVa}.

\begin{figure}
\begin{center}
\includegraphics[scale=0.6]{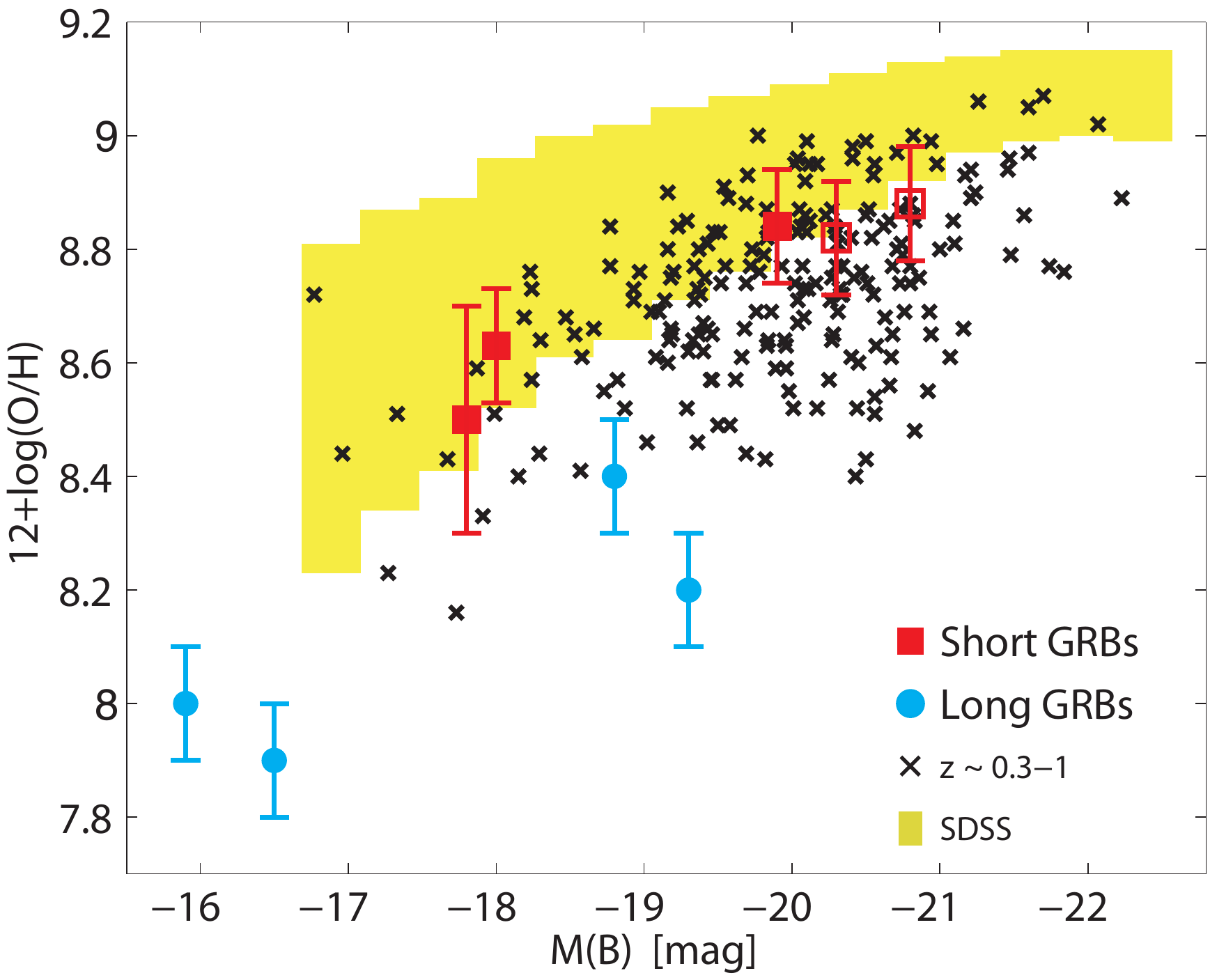}
\end{center}
\caption{\small\setlength{\baselineskip}{0.6\baselineskip}%
  Metallicity as a function of $B$-band absolute
  magnitude for the host galaxies of short (red) and long (blue)
  GRBs.  The yellow bars mark the $14-86$ percentile range for galaxies
  at $z\sim 0.1$ from the Sloan Digital Sky Survey \citep{thk+04},
  while crosses designate field galaxies at $z\sim 0.3-1$
  \citep{kk04}.
    Both field samples
  exhibit a clear luminosity-metallicity relation.  The long GRB hosts
  tend to exhibit lower than expected metallicities \citep{sgb+06},
  while the hosts of short GRBs have higher metallicities by about 0.6
  dex and are moreover in excellent agreement with the
  luminosity-metallicity relation.  From \cite{berger09}. }
\label{figIVc}
\end{figure}

This picture has not substantially changed in the \swift\ era.  To the
contrary, the result of various ground- and space-based efforts to
characterize GRB host galaxies (\citealt{lcf+06}, \citealt{cbc07},
\citealt{fps+08}, \citealt{sgl09}) have confirmed this basic outline
and expanded its domain of applicability to high redshift, combining
the power of the \swift\ burst catalog with {\it Spitzer}
observations.
At       higher redshifts, $z\simgt 3$, it seems particularly
interesting to use the GRB host galaxies to explore the evolution of
mass-metallicity relationships that are typically compiled using field
galaxies at low redshift and high-mass galaxy samples at high-redshift
(Figure~\ref{figIVc}).  Since GRB host redshifts are typically secured
via afterglow spectroscopy, the hosts themselves are uniquely free of
mass and luminosity selection effects.  In this context, even upper
limits provide useful constraints on mass-metallicity correlations and
their evolution with redshift (\citealt{cbc07}, \citealt{sgl09},
\citealt{berger09}).  Such studies have also served to place GRB host
galaxies in the context of other high-redshift galaxy populations
incuding the Lyman-break and Lyman-alpha emitting galaxies
\citep{fps+08}.

Finally, it is somewhat reassuring that numerical simulations of
star-forming galaxy populations can generate galaxy subpopulations
that reproduce the basic properties of the GRB host galaxies in
several important respects (\citealt{cbg04}, \citealt{cbf07}).  In
particular, selecting for high specific star-formation rate generates
mock galaxy catalogs with similar masses, luminosities, and colors to
GRB host galaxies.


\subsubsection{Metallicity matters.}
\label{subsub:hosts:long:metals}
While GRB host galaxies are often studied for the insights they
provide about larger astrophysical questions, including the history of
star formation through cosmic time, they also have the potential to
shed light on the nature of the GRB progenitors.  The association of
long-duration bursts with star formation, for example, was proposed
after observation of just two host galaxies
\citep{Paczynski:1998:GRB}, and demonstrated firmly from the
properties of the first twenty \citep{bkd02}.

Recent years have seen a surge of interest in the question of whether
GRB host galaxies,  and hence, presumably, GRB progenitors,  are
metal-poor by comparison to the larger population of star-forming
galaxies.  This question has been explored from a variety of
perspectives, and the result of these studies, still under active
debate, may eventually help refine our picture of the massive stellar
death behind each GRB.

The metallicities of GRB host galaxies, and indeed, detailed abundance
profiles, can be measured directly from high-resolution spectroscopy
of bright afterglows.  This area has seen dramatic progress in the
\swift\ era, with prompt arcsecond positions and visual magnitude
measurements from \swift\ UVOT and ground-based robotic telescopes
feeding rapid-response spectroscopy from large-aperture facilities.
One chief result of these efforts have been the collection of detailed
abundance characterizations for multiple bursts (\citealt{vel+04},
\citealt{cpb+05}, \citealt{fsl+06}, \citealt{2007ApJ_654_878B},
\citealt{pcb+07}, \citealt{cpr+07}, \citealt{psc+07},
\citealt{cpp+08}).  In addition, the metallicities of GRB host
galaxies have been measured via standard emission-line diagnostics
(\citealt{sgb+06}, \citealt{tfo+08}, \citealt{sgl09}), where possible,
and in some cases tentative conclusions have been drawn on the basis
of cruder relations such as mass-metallicity metrics \citep{bfk+07}
and host galaxy luminosities and morphologies \citep{fls+06}.

The conclusions of these studies have yet to be reconciled into a
single coherent picture of the nature of the GRB host galaxies and
their relationship to other low- and high-redshift galaxy populations.
However, the wealth of data do serve to define some associated issues
with reasonable clarity.  First, the metallicities of GRB host
galaxies at $z\simlt 1$ are significantly ($Z\sim 0.1 Z_\odot$)
sub-solar \citep{sgl09}, consistent with the sub-solar metallicities
measured for GRB host galaxies via absorption spectroscopy at $z\simgt
2$ (e.g., \citealt{cpp+08}).  These sub-solar metallicities are
neither surprising nor unusual for galaxy populations at high redshift
(\citealt{fps+08}, \citealt{sgl09}); moreover, several candidate
higher-metallicity hosts have been identified, although not yet
confirmed (\citealt{bfk+07}, \citealt{fls+06}).  At the same time, GRB
host galaxies seem to be readily distinguished, in luminosity and
morphology, from the host galaxies of core-collapse supernovae at
similar redshifts \citep{fls+06}, and the host galaxies of the
lowest-redshift $z\simlt 0.2$ bursts have uniformly low metallicities
that strongly distinguish them from the bulk of the low-redshift
galaxy population (\citealt{sgb+06}, \citealt{tfo+08}), and indeed,
from the host galaxies of nearby type Ibc supernovae \citep{mkk+08}.

These somewhat divergent findings might be reconciled in a picture
where the GRB progenitors prefer (or require) a low-metallicity
environment, since the increasing prevalence of such environments at
$z\simgt 1$ would allow GRB host galaxies to present an increasingly
fair sample of the population of star-forming galaxies at these higher
redshifts.  This argument would also dovetail with observations that
the GRB rate seems to increase with redshift faster than the cosmic
star formation rate, as mentioned above (\citealt{ld07},
\citealt{gp07}, \citealt{kyb+08}).  However, the claim that GRB host
galaxies represent a fair sample of star-forming galaxies, even at
$z\simgt 1$, remains in dispute (e.g., Figure~\ref{figIVc};
\citealt{berger09}).

A possibly significant implication of a metallicity-dependent GRB rate
would be an offset between the true star-formation rate and that
traced by GRBs.  If GRBs in low-metallicity environments and low-mass
galaxies are more luminous, then they are likely over-represented in
GRB samples.  Low-mass galaxies and galaxy outskirts have lower
metallicity on average and thus may yield more (and/or more luminous)
GRBs compared to high-mass galaxies \citep{rlb02}.  As galaxy mass
builds up through mergers, it is also possible that the highest-$z$
GRBs could be systematically more luminous due to their lower-mass
host galaxies, an intriguing hypothesis given the extreme luminosities
of some of the highest-redshift bursts of the \swift\ era, including
GRB\,050904.

Finally, it is worth noting that high-resolution afterglow spectra
show absorption features imprinted on the afterglow by gas at
multiple, widely divergent physical scales, possibly extending from
$d\sim 10$\,pc to Gpc distances (Section~\ref{subsub:hosts:long:subgal}).
With the GRB metallicity question ultimately referring to the nature
of the progenitor itself, searches for definitive signatures of the
progenitor's stellar wind material (\citealt{mfh+02},
\citealt{sgh+03}, \citealt{mhc+03}, \citealt{2007ApJ_654_878B},
\citealt{cpr+07}, \citealt{flv+08}, \citealt{pdr+08}) should continue
to be pursued and refined.  At the same time, in discussing the host
galaxies as a population, more common galaxy-integrated, emission-line
diagnostics may better serve to place GRB hosts in the proper
cosmological context.


\subsubsection{Subgalactic environments.}
\label{subsub:hosts:long:subgal}
\begin{figure}
\begin{center}
\includegraphics[width=5.2in]{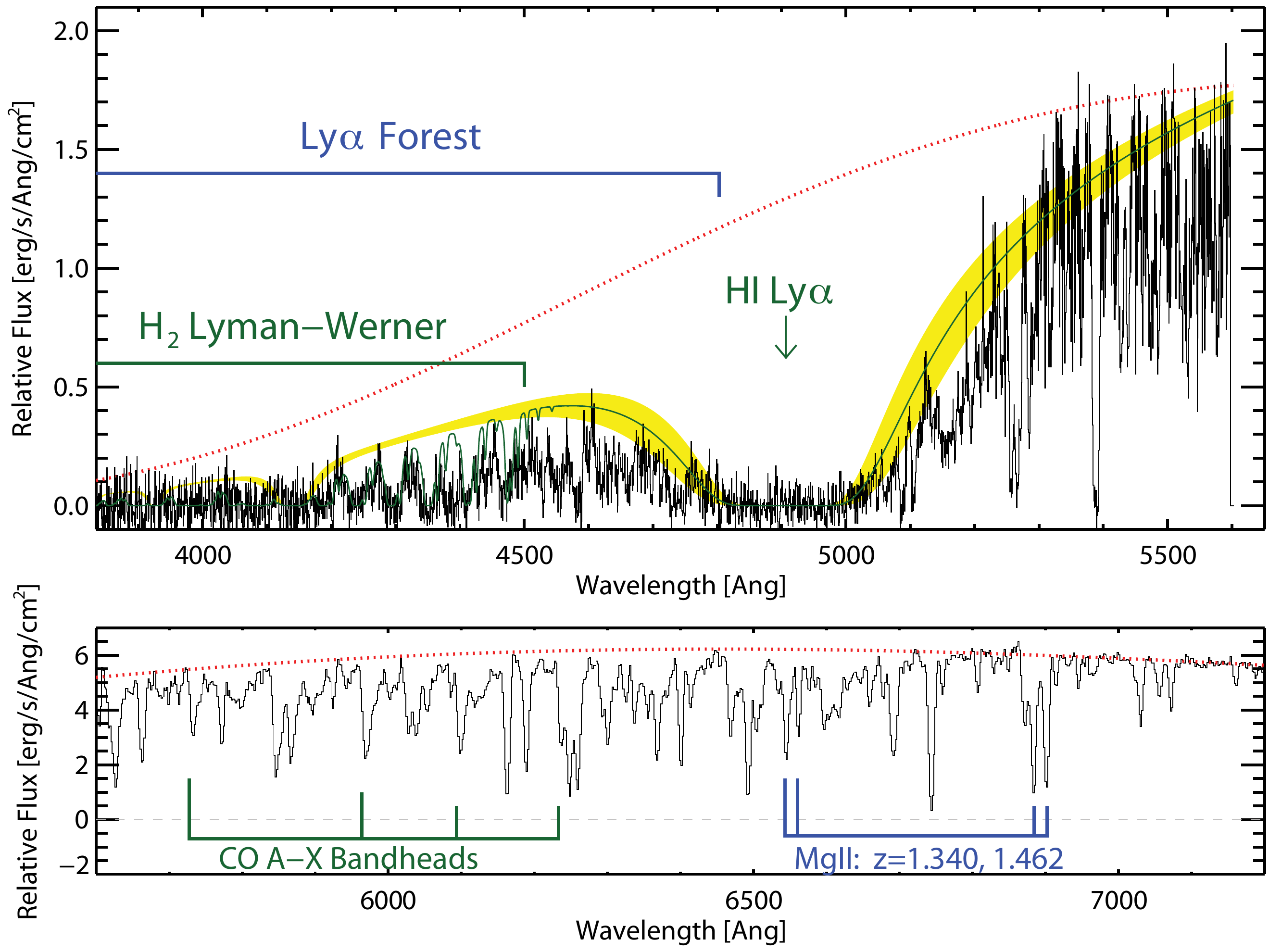}
\end{center}
\caption{\small
  Keck/LRIS spectrum of the afterglow for GRB~080607
  \citep{psp+09}.  The red dashed lines indicate a model of the
  intrinsic afterglow spectrum reddened heavily by dust in the host
  galaxy (rest-frame $A_V \approx 3.2$).  At $\lambda \approx
  4900$~\AA\ one identifies a damped \lya\ profile associated with
  \ion{H}{1} gas near the GRB.  The shaded region overplotted on the
  data corresponds to an \ion{H}{1} column density $\mnhi = 10^{22.7
    \pm 0.15} \cm{-2}$.  The model (green solid line) includes
  absorption from H$_2$ Lyman-Werner transitions.  The line opacity at
  $\lambda > 5500$~\AA\ is dominated by metal-line transitions from
  gas in the host galaxy and includes bandheads of the CO molecule.
  Suprisingly, this is the only sightline to date to show strong
  molecular absorption \citep{2007ApJ_668_667T}.  It also exhibits a
  roughly solar metallicity.  The figure also shows the spectral
  region or features corresponding to intergalactic Ly$\alpha$ and
  MgII absorption. The redshift for GRB 080607 is $z=3.036$.}
\label{figIVb}
\end{figure}

The same fast-response spectra that have enabled characterization of
elemental abundances in GRB host galaxies have led to a series of
discoveries regarding absorbing gas structures on sub-galactic scales
within the GRB host: a rich array of high-ionization fine-structure Fe
transitions in GRB\,051111 (\citealt{bpf+05}, \citealt{pbf+06},
\citealt{pcb06}); discovery of time-variability of such fine structure
features in three bursts, demonstrating excitation by UV photons from
the burst flash and young afterglow (\citealt{dcp+06},
\citealt{vls+07}, \citealt{dfp+08}); discovery of high-ionization
\ion{N}{5} features, providing evidence for absorption by gas within
$d\simlt 10$\,pc of the burst \citep{flv+08,pdr+08}; and most
recently, the first detection of molecular gas along the line of sight
to a GRB afterglow (Figure~\ref{figIVb}; \citealt{psp+09}).

The range of these discoveries reveals a surprising complication in
the interpretation of afterglow spectra. Depending on the burst, it
may be necessary to account for absorbing structures on every scale
from the immediate parsec-size circumburst environment, to surrounding
or intervening molecular clouds, to host galaxy absorbers on
kiloparsec scales, and ultimately, to low-ionization gas associated
with cosmological structures at gigaparsec distances.  Similar
complications are a feature of quasar absorption line studies; there,
however, the situation is simplified by the persistent nature and
overwhelming ionizing power of the quasar itself.

Indeed, the situation may be even more complicated than this.  The
discovery that the frequency ($dN/dz$) of strong \ion{Mg}{2} absorbers
along GRB lines of sight seems to be roughly four times the frequency
along quasar lines of sight \citep{ppc+06} suggests that some or most
such absorbers seen toward GRBs -- despite their large separation in
redshift from the host galaxy -- may be intrinsic to the GRB
environment \citep{pvl07}.


\subsection{Host Galaxies of Short Bursts}
\label{sub:hosts:short}

At the time of the \swift\ launch, the greatest mystery in GRB
astronomy was the origin of short GRBs.  A major step forward was made
in summer 2005 with the localization and afterglow detection of three
short bursts, GRB 050509B, 050709 and 070724.  These events were found
to be localized in regions of low star formation, either in low star
forming elliptical galaxies as for GRB 050509B
(\citealt{2005Natur_437_851G}, \citealt{2006ApJ_638_354B}) and GRB
050724 (\citealt{barthelmy05}, \citealt{bpc+05}) or in a region of a
galaxy with low star formation (\citealt{villasenor05},
\citealt{ffp+05}, \citealt{hwf+05}).  This was in stark contrast to
long bursts which are associated with star forming regions, and
implied a non massive core-collapse origin.

More than three years after these first short burst localizations, the
catalog of confidently-identified short burst host galaxies is growing
to the point where systematic studies can be carried out
\citep{berger09} -- although the effects of uncertain burst
attribution (i.e., long or short?), uncertain host identification
(especially for bursts with only \swift\ BAT or XRT localizations),
and unknown redshifts for faint candidate hosts conspire to keep any
conclusions largely qualitative at this time.

Indeed, without direct afterglow spectroscopy, association of short
bursts with candidate host galaxies and host galaxy clusters must be
approached probabilistically.  In cases where a well-localized
(preferably sub-arcsec) afterglow falls on a luminous region of the
candidate host, or within a high-mass or high-redshift cluster, the
association can probably be considered secure; however, in other cases
an a posteriori estimate of the probability of association must be
made \citep{2006AIPC_836_473B,fr07}.  Such estimates are inevitably
strongly dependent on input assumptions.  For example: What lifetime
and kick velocity distributions might be appropriate for progenitor
binary systems?  What about other possible progenitor classes?  Any
assumptions must be carefully considered before and after they are
applied.

With these caveats, a picture of the short burst host galaxy
population as a whole has developed (Figure~\ref{figIVa}). It consists
of three classes of host, two of which became apparent soon after the
short burst afterglow revolution of 2005: the low-redshift ($z\simlt
0.5$), high-mass ($L\sim L^*$), early-type host galaxies and galaxy
clusters, on the one hand (\citealt{2005Natur_437_851G},
\citealt{ffp+05}, \citealt{2006ApJ_638_354B}, \citealt{bpc+05},
\citealt{gno+08}), typified by the hosts of GRB\,050509B, GRB\,050724,
and GRB\,050813; and the low-redshift, sub-$L^*$, late-type galaxies
on the other, typified by the hosts of GRB\,050709
(\citealt{ffp+05},\citealt{cmi+06}), GRB\,051221 \cite{sbk+06}, and
GRB\,061006 \cite{dmc+09}.

The third class of short burst host galaxies consists of faint,
star-forming galaxies at $z\simgt 1$ (\citealt{bfp+07},
\citealt{cbn+08}, \citealt{berger09}), reminiscent of the host
galaxies of long bursts.  The existence of such higher-redshift,
star-forming short burst host galaxies was predicted  \citep{bpb+06}
based on binary population synthesis models assuming the compact
object merger model for the short bursts.  These simulations yield a
bimodal distribution of lifetimes for merging systems, with a spike of
mergers at short timescales, $\tau\simlt 100$\,Myr, followed by a
dominant merger population with a $\tau^{-1}$ lifetime distribution.

The association of short GRBs with both star-forming and early-type
galaxies has led to analogies with     type~Ia supernovae, whose host
demographics have similarly provided evidence for a wide distribution
of delay times between formation and explosion. At the same time, both
core-collapse supernovae and long-duration GRBs are observed (almost)
exclusively in late-type star-forming galaxies.  As with supernovae
and long-duration bursts, a detailed census of short burst redshifts,
host galaxy types, and burst locations within those hosts will
undoubtedly help to constrain progenitor models.


\subsection{Neither Long Nor Short}
\label{sub:hosts:neither}

Some interesting gamma-ray burst host galaxies are not obviously
associated with either the long or short burst classes.  In fact, the
``peculiar'' cases of GRB\,060505 and GRB\,060614 probably provide the
first examples where studies of the host galaxy properties have been
applied to argue for a (long- or short-) nature of the bursts
themselves \citep{fynbo06,gfp+06,ocg+07,tfo+08}.

GRB\,060614 was a low-redshift, long-duration burst with no detection
of a coincident supernova to deep limits. It was a bright burst
(fluence in $15-150$ keV band of 2.2 $\times$ 10$^{-5}$ erg cm$^{-2}$)
and well-studied in the X-ray and optical. With a T90 duration of
102\,s, it seems to fall squarely in the long burst category. A host
galaxy was found (\citealt{gfp+06}, \citealt{fynbo06},
\citealt{dellavalle06}) at $z=0.125$ and deep searches were made for a
coincident supernova. All other well-observed nearby GRBs have had
supernovae detected, but this one did not to limits $>$ 100 times
fainter than previous detections.

GRB\,060614 shares some characteristics with short bursts
\citep{gehrels06}.  The BAT light curve shows a first short,
hard-spectrum episode of emission (lasting 5 s) followed by an
extended and somewhat softer episode (lasting $\sim$100 s). The total
energy content of the second episode is 5 times that of the first.
Its light curve shape is similar in many respects to that of short
bursts with extended emission. 
      There are, however, differences in that the short
episode of this event is longer than the previously detected examples
and the soft episode is relatively brighter. Another similarity with
short bursts comes from a lag analysis of GRB\,060614. The lag between
temporal structures in the $50-100$ keV band and those in the $15-25$
keV bands for the first 5 s is 3 $\pm$ 6 ms which falls in the same
region of the lag-luminosity plot as short bursts
(Figure          5). It is difficult to determine unambiguously which
category of burst GRB\,060614 falls into. It is a long event by the
traditional definition, but it lacks an associated SN as had been
observed in all other nearby long GRBs. It shares some similarities
with \swift\ short bursts, but has important differences such the
brightness of the extended soft episode. If it is due to a collapsar,
it is the first indication that some massive star collapses either
fail as supernovae \citep{woosley93} or highly underproduce $^{56}$Ni
\citep{llr08}; if it is due to a merger, then the bright long-lived
soft episode is hard to explain within the framework of compact binary
mergers (\citealt{2002ApJ_579_706D}, \citealt{2004ApJ_608L_5L},
\citealt{2004MNRAS_352_753S}). Thus, this peculiar burst has
challenged the usual classifications of GRBs.


\section{BASIC PHYSICAL CONSIDERATIONS}
\label{sec:theory}

In this section, we endeavor to outline some of the physical processes
that are believed to be most relevant to interpreting GRBs.  Though
the field is far from maturity, sufficient progress has been made
in identifying the essential physical ingredients.  A basic scheme can
provide a conceptual framework for describing the observations even
when the framework is inaccurate! The following should be interpreted in
this spirit.

\begin{figure}
\flushleft\hspace*{-0.8in}\includegraphics[width=6.15in]{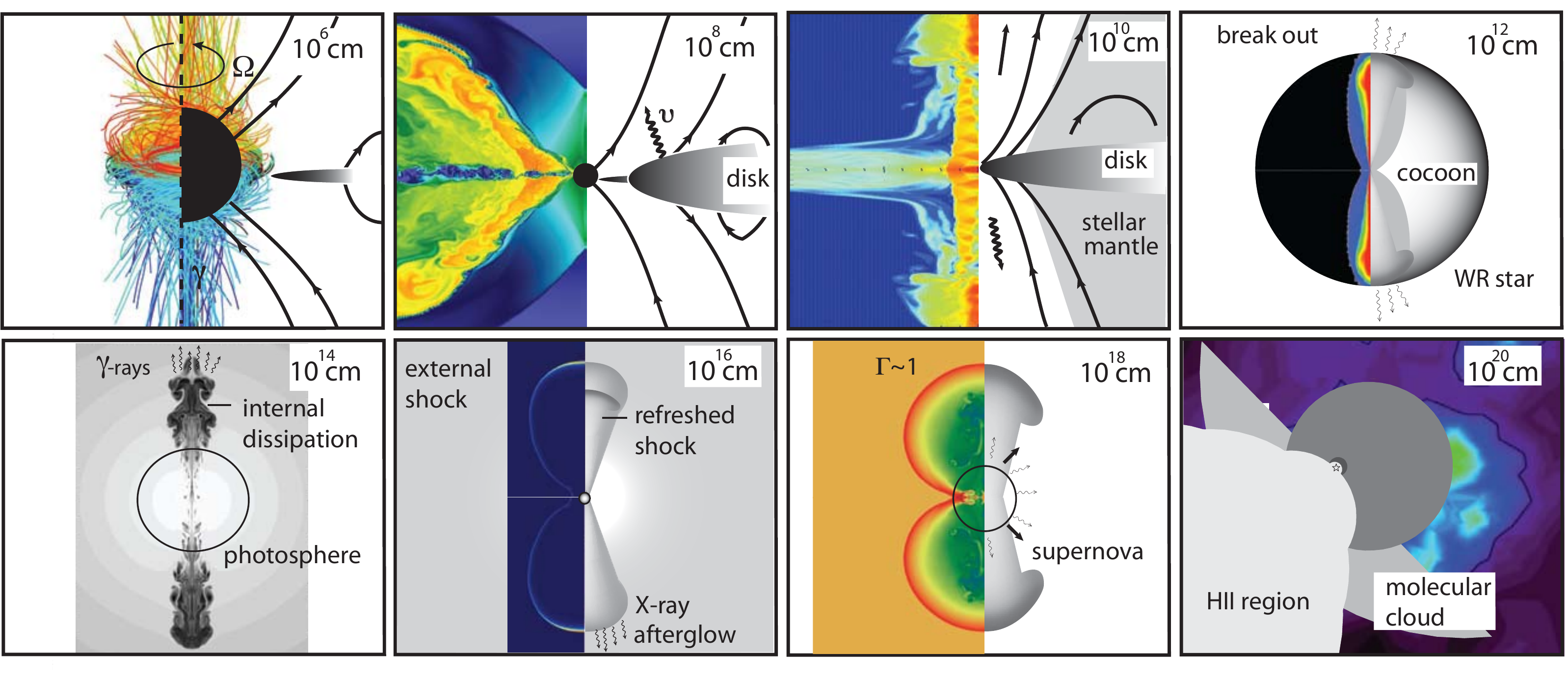}
\caption{\small
  Diagram exhibiting GRB activity over successive
  decades in radius ranging from $10^6$ cm to 1pc. The first task in
  attempting to construct a general scheme of GRBs is to decide which
  parameters exert a controlling influence upon their properties. 
  ($10^6$\,cm) The GRB nucleus (BH or NS) and its
  magnetosphere.
  ($10^8$\,cm) The accretion flow is likely to be embedded in a very
  active corona. We expect coronal arches, as well as large magnetic
  structures, to be quite common and to be regenerated on an orbital
  timescale.  Relativistic outflow from the black hole is proposed to
  be focused into two jets. 
  ($10^{10}$\,cm) Even if the outflow is not narrowly collimated, some
  beaming is expected because energy would be channelled
  preferentially along the rotation axis. The majority of stellar
  progenitors, with the exception of some very compact stars, will not
  collapse entirely during the typical duration of a GRB. A stellar
  envelope will thus remain to impede the advance of the jet.
  ($10^{12}$\,cm) This is the typical size of an evolved massive star
  progenitor.  A thermal break-out signal should precede the
  canonical, softer $\gamma$-rays observed in GRBs.  
  ($10^{14}$\,cm) Velocity differences across the jet profile provide
  a source of free energy.  The most favourable region for shocks
  producing highly variable $\gamma$-ray light curves is above the
  baryonic or pair-dominated photosphere. 
  ($10^{16}$\,cm) The external shock becomes important when the
  inertia of the swept-up external matter starts to produce an
  appreciable slowing down of the ejecta. 
  ($10^{18}$\,cm) Finally, we come to the end of the relativistic
  phase. This happens when the mass $E/c^2$ has been swept-up.  }
\label{figVa}
\end{figure}

GRB activity manifests itself over a dynamical range of $\sim$ 13
decades in radius. In Figure \ref{figVa}, we show a schematic montage
of successive decades, exhibiting phenomena which are believed to take
place on each of these length scales. The phenomena are not directly
observed and the associated frames represent educated guesses of their
geometrical arrangements.  An {\it anatomical} summary of the
underlying physical processes working outward from the smallest to the
largest scales follows.

\subsection{The Central Engine}

In principle, flow onto a compact object can liberate gravitational
potential energy at a rate approaching a few tenths of $\dot{M}c^2$,
where $\dot{M}$ is the mass inflow rate. Even for such high
efficiencies the mass requirements are rather large, with the more
powerful GRB sources $\sim 10^{53}\;{\rm erg\;s}^{-1}$ having to
process upwards of $10^{-2}M_\odot\;{\rm s}^{-1}$ through a region
which is not much larger than the size of a NS or a stellar mass
BH. Radiation of the BH rest mass on a timescale $r_{\rm g}/c$, where
$r_{\rm g}=GM/c^2=1.5 \times 10^5 (M/M_\odot)$ cm is the
characteristic size of the collapsed object, would yield luminosities
$c^5/G=4\times 10^{59}$ erg s$^{-1}$.

When mass accretes onto a BH or NS under these conditions, the
densities and temperatures are so large that photons are completely
trapped and neutrinos, being copiously emitted, are the main source of
cooling.  The associated interaction cross section is then many orders
of magnitude smaller, and, as a result, the allowed accretion rates
and luminosities are correspondingly much higher. For example, using
the cross section for neutrino pair production\footnote{Although we
  have considered here the specific case of neutrino pair creation,
  the estimates vary little when one considers, for example, coherent
  scattering of neutrinos by nuclei and/or free nucleons (except for
  the energy scaling).}, the Eddington limit can be rewritten as
$L_{{\rm Edd},\nu}=8 \times 10^{53} ({E_\nu/50 {\rm MeV}})^{-2}
(M/M_{\odot})\; {\rm erg~s^{-1}}$, with an associated accretion rate
is $\dot{M}_{{\rm Edd},\nu}\times\;{\rm (efficiency)}^{-1}$, where
$\dot{M}_{{\rm Edd},\nu}=L_{{\rm Edd},\nu}/c$
(\citealt{rr06}, 
 \citealt{lr07}).

The blackbody temperature if a luminosity $L_{{\rm Edd},\nu}$ emerges
from a sphere of radius\footnote{Similar overall fiducial numbers also
  hold for neutron stars, except that the simple mass scalings
  obtained here are lost.} $r_{\rm g}$ is
\begin{equation}
T_{{\rm Edd},\nu}=\left({L_{{\rm Edd},\nu} \over 4\pi r_{\rm g}^2
  \sigma_{\rm SB}}\right)^{1/4}\ \sim 45 \;(M/M_{\odot})^{-1/4}
\left({E_\nu \over 50 {\rm MeV}}\right)^{-1/2}\;{\rm MeV}.
\end{equation}
The radiation temperature is expected to be $\leq T_{\rm th}=G M
m_{\rm p}/ (3k r_{\rm g})\sim 200$ MeV, the temperature the accreted
material would reach if its gravitational potential energy were turned
entirely into thermal energy.

Related to this, there is a fiducial density in the vicinity of the BH
\begin{equation}
\rho_{{\rm Edd},\nu} = {\dot{M}_{{\rm Edd},\nu} \over 4\pi r_{\rm
    g}^{2} c} \sim 10^{11} (M/M_{\odot})^{-1} \left({E_\nu \over 50
  {\rm MeV}}\right)^{-2} \;{\rm g~cm^{-3}}.
\end{equation}
It should be noted that the typical Thomson optical depth under these
conditions is $\tau_{\rm T} \sim n_{{\rm Edd},\nu}^{1/3} r_{\rm g}
\sim 10^{16}$ and so, as expected, photons are incapable of escaping
and constitute part of the fluid. A characteristic magnetic field
strength is that for which $B^2/8\pi=n_{{\rm Edd},\nu}m_p c^2$:
\begin{equation}
B_{{\rm Edd},\nu} = \left({L_{{\rm Edd},\nu} \over R_{\rm
    g}^2c}\right)^{1/2}\sim 3\times 10^{16} (M/M_{\odot})^{11/2}
\left({E_\nu \over 50 {\rm MeV}}\right)^{-1} \;{\rm G}.
\end{equation}
If a field $B_{{\rm Edd},\nu}$ were applied to a BH with $J\approx
J_{\rm max}$, the electromagnetic power extraction would be $\sim
L_{{\rm Edd},\nu}$ \citep{1992Natur_357_472U}.

\subsection{Accretion Flows}

As we discussed above, a BH or NS embedded in infalling matter offers
a more efficient power source than any other conceivable
progenitor. Although this efficiency is over a hundred times larger
than that traditionally associated with thermonuclear reactions, the
required rate of mass supply for a typical GRB is of course
colossal. Such high mass fueling rates are never reached for BHs in
XRBs or AGN, where the luminosity remains well below the {\it photon}
Eddington limit. They can, however, be achieved in the process of
forming NS and stellar--mass BHs during the collapse of massive
stellar cores (\citealt{hc91}, \citealt{woosley93},
\citealt{macfadyen99}, \citealt{npk01}) and in binary mergers involving
compact objects
(\citealt{rj97}, 
 \citealt{kl98}, 
 \citealt{rosswog2003}, 
 \citealt{lrp05}, 
 \citealt{srj06},
 \citealt{mpq08},
 \citealt{2004MNRAS.351.1121R}).
Consequently, most recent theoretical work has been directed towards
describing the possible formation channels for these systems
(\citealt{fwh99},
 \citealt{bbr02},
 \citealt{irt04},
 \citealt{2004ApJ...607L..17P}). 
This involves evaluating those which are likely to
produce a viable central engine, and understanding the flow patterns
near relativistic objects accreting matter in the hypercritical regime
(\citealt{2002ApJ_579_706D},
 \citealt{cb06},
 \citealt{pz06}, 
 \citealt{paz06}, 
 \citealt{sst07}, 
 \citealt{knj08}),
where photons are unable to provide cooling but neutrinos do so
efficiently.
 
The angular momentum is quite generally a crucial parameter, in many
ways determining the geometry of the flow \citep{lrr06}.  The
quasi-spherical approximation breaks down when the gas reaches a
radius $r_{\rm circ} \sim l^2/GM$, where $l$ is the angular momentum
per unit mass, and if injection occurs more or less isotropically at
large radii, a familiar accretion disk will form. The inner regions of
disks with mass fluxes $\leq \dot{M}_{{\rm Edd},\nu}$ are generally
able to cool by neutrinos on time scales shorter than the inflow time.

If $\dot{m}=\dot{M}/\dot{M}_{{\rm Edd},\nu} \leq 1$, then the bulk of
the neutrino radiation comes from a region only a few gravitational
radii in size, and the physical conditions can be scaled in terms of
the {\it Eddington quantities} defined above. The remaining relevant
parameter, related to the angular momentum, is $v_{\rm inflow}/v_{\rm
  freefall}$, where $v_{\rm freefall}\simeq(2GM/R)^{1/2}$ is the free
fall velocity. The inward drift speed $v_{\rm inflow}$ would be of
order $v_{\rm freefall}$ for supersonic radial accretion. When angular
momentum is important, this ratio depends on the mechanism for its
transport through the disk, which is related to the effective shear
viscosity.  For a thin disk, the factor $(v_{\rm inflow}/v_{\rm
  freefall})$ is of order $\alpha (H/R)^{2}$, where $H$ is the scale
height at radius $R$ and $\alpha$ is the phenomenological viscosity
parameter \citep{ss73}.

The characteristic density, at a distance $r$ from the hole, with
account of the effects of rotation, is
\begin{equation}
\rho\sim \dot{m}(r/r_{\rm g})^{-3/2} (v_{\rm inflow}/v_{\rm
  freefall})\rho_{{\rm Edd},\nu},
\end{equation}
and the maximum magnetic field, corresponding to equipartition with
the bulk kinetic energy, would be
\begin{equation}
B_{\rm eq}\sim \dot{m}^{1/2}(r/r_{\rm g})^{-5/4}(v_{\rm inflow}/v_{\rm
  freefall})^{1/2}B_{{\rm Edd},\nu}.
\end{equation}
Any neutrinos emerging directly from the central core would have
energies of a few MeV. Note that, as mentioned above, $kT_{{\rm
    Edd},\nu}$ is far below the virial temperature $kT_{\rm vir}\simeq
m_p c^2(r/r_{\rm g})$.  The flow pattern when accretion occurs would
be then determined by the value of the parameters $L_\nu/L_{{\rm
    Edd},\nu}$, which determine the importance of radiation pressure
and gravity, and the ratio $t_{\rm cool}/t_{\rm dynamical}$, which
fixes the temperature if a stationary flow pattern is set up.

It is widely believed that accretion onto a compact object, be it a NS
or a stellar mass BH, offers the best hope of understanding the ``prime
mover'' in all types of GRB sources although a possible attractive
exception includes a rapidly spinning neutron star with a powerful
magnetic field
(\citealt{1992Natur_357_472U},
 \citealt{t94},
 \citealt{s99},
 \citealt{tcq04},
 \citealt{um07},
 \citealt{bqam08},
 \citealt{dblo08}). If
there is an ordered field $B$, and a characteristic angular velocity
$\omega$, for a spinning compact source of radius $r_\ast$, then the
magnetic dipole moment is $\sim B r_\ast^3$. General arguments suggest
\citep{pacini67,gold68} that the non-thermal magnetic-dipole-like
luminosity will be $L_{\rm em} \approx B^2 r_\ast^6 \omega^4/c^3$, and
simple scaling from these familiar results of pulsar theory require
fields of order $10^{15}$~G to carry away the rotational or
gravitational energy (which is $\sim 10^{53}$~erg) in a time scale of
seconds \citep{1992Natur_357_472U}.

\subsection{Jet Production}

There are two ingredients necessary for the production of jets. First,
there must be a source of material with sufficient free energy to
escape the gravitational field of the compact object. Second, there
must be a way of imparting some directionality to the escaping
flow. Our eventual aim must be to understand the overall flow pattern
around a central compact object, involving accretion, rotation, and
directional outflow but we are still far from achieving this.  Most
current workers who have discussed outflow and collimation have simply
invoked some central supply of energy and material. A self-consistent
model incorporating outflow and inflow must explain why some fraction
of the matter can acquire a disproportionate share of energy (i.e., a
high enthalpy).

A spinning black hole (or neutron star) constitutes an excellent
gyroscope, and the ingredients of accretion, angular momentum, entropy
production, and possibly magnetic fields are probably sufficient to
ensure the production of collimated outflow
\citep{mckinney06}. However, the detailed mechanism is a matter for
debate (it is not even clear what is accelerated), and several
distinct lines of research are currently being pursued. One solution
is to reconvert some of this energy via collisions outside the disk
into electron-positron pairs or photons
(\citealt{mochkovitch93},
 \citealt{gdn87},
 \citealt{rosswogrr02},
 \citealt{rrrd03},
 \citealt{aloy05},
 \citealt{dobrl09}).
If this occurs in a region of low baryon density (e.g., along the
rotation axis, away from the equatorial plane of the disk) a
relativistic pair-dominated wind can be produced. An obvious
requirement for this mechanism to be efficient is that the neutrinos
escape (free streaming, or diffusing out if the density is high
enough) in a time scale shorter than that for advection into the
BH. The efficiency for conversion into pairs (scaling with the square
of the neutrino density) is too low if the neutrino production is too
gradual, so this can become a delicate balancing act.

Jets may alternatively be produced electromagnetically. Such mechanism
could, in principle, circumvent the above restriction in
efficiency. The potential difference across a disk threaded by an open
magnetic field lines can exceed $10^{22} V$, and this is available for
accelerating high-energy particles which will produce an
electron-positron cascade and ultimately a relativistic jet that
carries away the binding energy of the accreting gas.  Blandford \&
Znajek (1977) extended this idea to black holes and showed how the
spin energy of the hole could likewise be extracted. A hydromagnetic
description of this mechanism is more likely to be appropriate
\citep{bp82}. The field required to produce $L_{\rm em} \geq
10^{51}\;{\rm erg\;s^{-1}}$ is enormous, and may be provided by a
helical dynamo operating in hot, convective nuclear matter with a
millisecond period \citep{duncan92}.  A dipole field of the order of
$10^{15}$~G appears weak compared to the strongest field
\citep{tqb05,price06} that can in principle be generated by
differential rotation ($\sim 10^{17}[P/1\;{\rm ms}]^{-1}\;{\rm G}$),
or by convection ($\sim 10^{16}\;{\rm G}$), although how this may come
about in detail is not resolved. Note, however, that it only takes a
residual torus (or even a cold disk) of $10^{-3}\ M_\odot$ to confine
a field of $10^{15}$~G.

A potential death-trap for such relativistic outflows is the amount of
entrained baryonic mass from the surrounding medium. For instance, a
Poynting flux of $10^{53}$ erg could not accelerate an outflow to
$\Gamma \geq 10^2$ if it had to drag more than $\sim 10^{-4}M_\odot$
of baryons with it. A related complication renders the production of
relativistic jets even more challenging, because the high neutrino
fluxes are capable of ablating baryonic material from the surface of
the disk. Thus a rest mass flux $\dot{M}_\eta$ limits the bulk Lorentz
factor of the wind to $\Gamma_\eta=L_{\rm wind}/(\dot{M}_\eta c^2)$.
Assuming that the external poloidal field strength is limited by the
vigor of the convective motions, the spin-down luminosity scales with
neutrino flux as $L_{\rm wind} \approx L_{\rm em}\propto B^2\propto
v_{\rm con}^2\propto L_\nu^{2/3}$, where $v_{\rm con}$ is the
convective velocity.  The ablation rate is $\propto L_\nu^{5/3}$
\citep{qw96,mtq07}, which indicates that the limiting bulk Lorentz
factor $\Gamma_\eta$ of the wind decreases as $L_\nu^{-1}$. Thus the
burst luminosity emitted by a magnetized neutrino cooled disk may be
self-limiting. Mass loss could, however, be suppressed if the
relativistic wind were somehow collimated into a jet. This suggests
that centrifugally-driven mass loss will be heaviest in the outer
parts of the disk, and that a detectable burst may be emitted only
within a relatively small solid angle centered on the rotation axis
\citep{le00}.

\subsection{Jet Collimation, Stability, and Confinement}

As we discussed above, one of the key issues one needs to address
concerning jets is how they are formed? A second one is, how do they
retain their coherence and collimation as they transverse circumburst
space? The second issue is no less pressing than the first, since
jet-like flows known on Earth are notoriously unstable.  The situation
in a jet is nonetheless different from that encountered in the
laboratory because of the super-Alfv\'enic and supersonic streaming
velocity \citep{mb08}.  MHD probably provides a better description of
the macroscopic behavior of a GRB jet than in the case of laboratory
plasmas because the particle Larmor radii are so much smaller than the
transverse size of the jet. Formal stability analysis of even the
simplest jet models is complex. Unstable MHD modes (such as pinch and
kink instability) do not grow as rapidly as in stationary plasma,
although sufficiently short-wavelength unstable modes, localized
within the jet interior are not suppressed by the relative
motion. Longitudinal magnetic field in the core of the jet can
likewise act as a backbone, provided that the correlation length of
the magnetic field reversal along the jet exceeds the wavelength of
the perturbation. Instabilities may be suppressed for perturbations of
large wavelength along the jet by the inertia of the ambient medium.
Whatever one's view of the relative merits of fluid and
electromagnetic models of relativistic jets -- and perhaps the truth
lies between the two extremes -- it is clear that our understanding of
the role of magnetic fields in jets is less advanced than that of
other aspects of the problem.

An understanding of the collimation and confinement of a jet can come
about only through a knowledge of the properties of the medium through
which it propagates. Information about the ambient pressure gradient
propagates into the jet at the internal sound speed. If the jet moves
through a background with pressure scale height $\sim r$, the
necessary condition for the jet interior to remain in pressure balance
with its surroundings is $\theta M_j \leq 2$, where $M_j$ is the Mach
number and $\theta\sim d/r$ is the opening angle. If this condition is
not satisfied when a jet is moving into a region of higher pressure,
then strong shocks are driven into the jet. A jet moving into a region
of lower pressure becomes overpressured relative to its surroundings,
and thereafter expands freely.

Freedom or confinement? This is the first question for which
an extrinsic or environmental effect comes into play
(which may, in turn, strongly affect what we observe),
and it is
thought to be particularly important for massive stellar GRB
progenitors. This is because in such stars, a stellar envelope will
remain to impede the advance of the jet
(\citealt{macfadyen99}, 
 \citealt{mwh01},
 \citealt{mre01},
 \citealt{aloy00},
 \citealt{rrcr02},
 \citealt{mat03},
 \citealt{zwh04},
 \citealt{2003ApJ...599L...5P}). 
The beam will then evacuated a channel out to some location where it
impinges on the stellar envelope. A continuous flow of relativistic
fluid emanating from the nucleus supplies this region with mass,
momentum, energy, and magnetic flux.  Most of the energy output during
that period is deposited into a cocoon or {\it wastebasket}
surrounding the jet, which, after expansion, would have enough kinetic
energy content to substantially alter the structure of the
relativistic outflow, if not, in fact, provide much of the observed
explosive power \citep{rrcr02}.

Can relativistic jets really be formed inside stars?  This is not the
most propitious environment to create an ultra relativistic,
baryons-starved jet. What needs to be demonstrated, is that the
outflow is not ``poisoned'' by baryons by the time that it reaches the
surface of the star.  It appears that it is not necessary to collimate
the jet very tightly or to achieve high bulk Lorentz factor as the
flow leaves the stellar surface (\citealt{zwh04},
                                 \citealt{mlb07}).  As long as the
emergent flow has a high enthalpy per baryon, it will expand and
achieve its high terminal speed some distance from the star
(\citealt{um07},
 \citealt{bqam08}). 
  A strong thermal break-out signal is expected to
precede the canonical $\gamma$-rays observed in GRBs with massive
progenitors as the shock breaks through the stellar surface and
exposes the hot shocked material
(\citealt{mwh01},
 \citealt{rml02},
 \citealt{wm03},
 \citealt{wmc07}). For
very extended envelopes, the jet may be unable to break through the
envelope. TeV neutrino signals produced by Fermi accelerated
relativistic protons within the cork may provide a means of detecting
such choked-off, $\gamma$-ray dark collapses \citep{mw01}.

\subsection{Dissipation and Cooling Effects Within the Jet}

The unique feature of GRBs is that the bulk Lorentz factor $\Gamma$
may reach values from hundreds to thousands
\citep{2001ApJ_555_540L}. The relativistic motion of the radiating
particles introduces many interesting effects in GRB emissions that
must be properly taken into account (e.g. \citealt{ed96}).

Three frames of reference are considered when discussing the emission
from systems moving with relativistic speeds: the stationary frame,
which is denoted here by asterisks, the comoving frame, denoted by
primes, and the observer frame. The differential distance traveled by
the expanding source during differential time $dt_*$ is $dr = \beta c
dt_*$, where $\beta = \sqrt{1-\Gamma^{-2}}$. Due to time dilation, $dr
= \beta \Gamma c dt^\prime$. The relationship between comoving and
observer times is $(1+z)\Gamma dt^\prime (1-\beta\cos\theta) =
(1+z)dt^\prime/\delta = dt$, where $\theta$ is the angle between the
emitting element and the observer, $\delta =
[\Gamma(1-\beta\cos\theta)]^{-1}$ is the Doppler factor and $z$ the
cosmological redshift. For an on-axis observer we therefore see that
$dt \cong (1+z)dr /\Gamma^2 c$, and, as a result, the blast wave can
travel a large distance $\Gamma^2 c \Delta t$ during a small observing
time interval. A photon detected with dimensionless energy $\epsilon =
h\nu/m_ec^2$ is emitted with energy $\delta\epsilon^\prime/(1+z)$.

Few would dispute the statement that the photons which bring us all
our information about the nature of GRBs are the result of
triboluminescence. For instance, velocity differences across the jet
profile provide a source of free energy from particle acceleration
through shock waves, hydromagnetic turbulence, and tearing mode
magnetic reconnection \citep{1994ApJ_430L_93R}.  If the value of
$\Gamma$ at the base increases by a factor $\geq 2$ over a timescale
$\Delta t$, then the later ejecta will catch up and dissipate a
fraction of their energy at radius given by
\begin{equation}
r_{\iota} \sim c \Delta t \Gamma^2 \sim 3 \times 10^{14} \Delta t_0
\Gamma_2^2~{\rm cm},
\end{equation}
where $\Delta t= 1 \Delta t_0$ s and $\Gamma=10^2\Gamma_2$.
Dissipation, to be most effective, must occur when the wind is
optically thin: $\tau_T \simeq n'\sigma_T (r/\Gamma)\leq 1$ (here $
n'$ is the comoving number density).  Otherwise it will suffer
adiabatic cooling before escaping \citep{1986ApJ_308L_47G}. The
photosphere (baryonic or pair-dominated) is a source of soft
thermalized radiation, which may be observationally detectable in some
GRB spectra and may also result in inverse Compton cooling of the
nonthermal electrons accelerated in the shocks occurring outside it
(\citealt{mre00},
 \citealt{rll02},
 \citealt{spm00},
 \citealt{krm02},
 \citealt{2002ApJ...578..812M},
 \citealt{rr2005},
 \citealt{gs2007},
 \citealt{pmr06},
 \citealt{rp08},
 \citealt{tmr07}).

In the presence of turbulent magnetic fields built up behind the
shocks, the electrons can produce a synchrotron power-law radiation
spectrum, whereas the inverse Compton scattering of these synchrotron
photons extends the spectrum into the GeV range
\citep{1994ApJ_432_181M}. To illustrate the basic idea, suppose that
electrons, protons, and magnetic field share the available internal
energy, then the electrons reach typical random Lorentz factors of
$\gamma \sim m_p/m_e$; while the assumption of a Poynting flux $L_B$
implies a comoving magnetic field of order $B \sim L_B^{1/2}
r_\iota^{-1} \Gamma^{-1} \sim 10^{5}L_{B,50}^{1/2}r_{\iota,13}^{-1}
\Gamma_2^{-1}\; {\rm G}$, where $L_B= 10^{50}L_{B,50}$ erg s$^{-1}$
and $r_{\iota} =10^{13} r_{\iota,13}$. For these values of $\gamma$
and $B$, the typical observed synchrotron frequency is $\nu_{\rm sy}
\sim 0.5 L_{B,50}^{1/2} r_{\iota,13}^{-1} (1+z)^{-1}$ MeV, independent
of the bulk Lorentz factor $\Gamma$, and in excellent agreement with
the observed values of the $\nu F_\nu$ peak of GRB spectra (Figure
1). Yet, there are in some instances serious problems associated with
this model (e.g., dissipation efficiencies). These difficulties have
motivated consideration for alternative scenarios (e.g.,
\citealt{2008arXiv0812.0021K}, \citealt{2008MNRAS.384...33K}).

A magnetic field can ensure efficient cooling even if it is not strong
enough to be dynamically significant. If, however, the field is
dynamically significant in the wind \citep{mre1997}, then its internal
motions could lead to dissipation even in a constant velocity wind
\citep{t94}.  Instabilities in this magnetized wind may be responsible
for particle acceleration \citep{th06}, and it is possible that
$\gamma-$ray production occurs mainly at large distances from the source
\citep{lyutikov03}.

A further effect renders the task of simulating unsteady winds even
more challenging. This stems from the likelihood that any entrained
matter would be a mixture of protons and neutrons. If a streaming
velocity builds up between ions and neutrons then interactions can
lead to dissipation even in a steady jet where there are no shocks
(\citealt{derishev99},
 \citealt{bel03}).

\subsection{Jet Interaction with the External Environment}

Astrophysicists understand supernova remnants reasonably well, despite
continuing uncertainty about the initiating explosion; likewise, we
may hope to understand the afterglows of GRBs, despite the
uncertainties about the trigger that we have already emphasized.

In the simplest version of the afterglow model, the blast wave is
approximated by a uniform thin shell. A forward shock is formed when
the expanding shell accelerates the external medium, and a reverse
shock is formed due to deceleration of the cold shell. The forward and
reverse shocked fluids are separated by a contact discontinuity and
have equal kinetic energy densities.

As the blast wave expands, it sweeps up material from the surrounding
medium to form an external shock \citep{mr93}. Protons captured by the
expanding blast wave from the external medium will have total energy
$\Gamma m_p c^2$ in the fluid frame, where $m_p$ is the proton
mass. The kinetic energy swept into the comoving frame by an
uncollimated blast wave at the forward shock per unit time is given by
\citep{bm76} $dE^\prime/dt^\prime = 4\pi r^2 n_{\rm ext} m_p c^3
\beta\Gamma(\Gamma-1)$, where factor of $\Gamma$ represents the
increase of external medium density due to length contraction, the
factor ($\Gamma -1$) is proportional to the kinetic energy of the
swept up particles, and the factor $\beta$ is proportional to the rate
at which the particles are swept.

The external shock becomes important when the inertia of the swept-up
external matter starts to produce an appreciable slowing down of the
ejecta. The expanding shell will therefore begin to decelerate when $E
= \Gamma M_{\rm b} c^2 = \Gamma^2 m_p c^2 (4\pi r_{\rm d}^3 n_{\rm
  ext}/3) $, giving the deceleration radius
(\citealt{mr93},
 \citealt{1992MNRAS_258P_41R})
\begin{equation} 
r_{\rm d} = \left( {3 E \over 4\pi \Gamma^2 c^2 m_p n_{\rm
    ext}}\right)^{1/3} \sim 3 \times 10^{16} \left({E_{52}\over
  \Gamma_{2}^2 n_{\rm ext}}\right)^{1/3}\;\rm{cm},
\end{equation}
 where $\Gamma \cong E/M_{\rm b} c^2$ is the coasting Lorentz factor,
$M_{\rm b}$ is the baryonic mass, $E= 10^{52}E_{52}$ erg is the
apparent isotropic energy release and $n_{\rm ext}$ is the number
density of the circumburst medium.  This sets a characteristic
deceleration length. This deceleration allows slower ejecta to catch
up, replenishing and re-energizing the reverse shock and boosting the
momentum in the blast wave.

Most treatments employing blast-wave theory to explain the observed
afterglow emission from GRBs assume that the radiating particles are
electrons. The problem here is that $\sim m_p/m_e \sim 2000$ of the
nonthermal particle energy swept into the blast-wave shock is in the
form of protons or ions, unless the surroundings are composed
primarily of electron-positron pairs. For a radiatively efficient
system, physical processes must therefore transfer a large fraction of
the swept-up energy to the electron component
\citep{2008PhRvE..77b6403G}. In most elementary treatments it is
simply assumed that a fraction $\epsilon_e$ of the forward-shock power
is transferred to the electrons.

The strength of the magnetic field is another major uncertainty. The
standard prescription is to assume that the magnetic field energy
density $u_B$ is a fixed fraction $\epsilon_B$ of the downstream
energy density of the shocked fluid. Hence $u_B=B^2/(8\pi) =
4\epsilon_B n_{\rm ext} m_p c^2(\Gamma^2-\Gamma)$ (although see, e.g.,
\citealt{2003MNRAS.339..881R}).  It is also generally supposed in
  simple blast-wave model calculations that some mechanism injects
  electrons with a power-law distribution between electron Lorentz
  factors $\gamma_{\rm min} \leq \gamma \leq \gamma_{\rm max}$
  downstream of the shock front, where the maximum injection energy is
  obtained by balancing synchrotron losses and an acceleration rate
  given in terms of the inverse of the Larmor time scale.

A break is formed in the electron spectrum at cooling electron Lorentz
factor $\gamma_{\rm c}$, which is found by balancing the synchrotron
loss time scale $t_{\rm sy}^\prime$ with the adiabatic expansion time
$t^\prime_{\rm adi} \sim r/(\Gamma c) $ \citep{sari1998}. For an
adiabatic blast wave, $\Gamma\propto t^{-3/8}$, so that $\gamma_{\rm
  min} \propto t^{-3/8}$ and $\gamma_{\rm c} \propto t^{1/8}$. As a
consequence, the accelerated electron minimum random Lorentz factor
and the turbulent magnetic field also decrease as inverse power-laws
in time (\citealt{mr93},
         \citealt{1992MNRAS_258P_41R}). This implies that the
spectrum softens in time, as the synchrotron peak corresponding to the
minimum Lorentz factor and field decreases, leading to the possibility
of late long wavelength emission \citep{sari1998}.
 
The relativistic expansion is then gradually slowed down, and the
blast wave evolves in a self-similar manner with a power-law
lightcurve. This phase ends when so much mass shares the energy that
the Lorentz factor drops to 1 \citep{ap01,rrm08}. Obviously, this
happens when a mass $E/c^2$ has been swept up. This sets a
non-relativistic mass scale:
\begin{equation}
\label{l_S}
r_{\rm s} = \Gamma^{2/3} r_{\rm d} = \left( {3 E\over 4\pi m_p c^2
  n_{\rm ext}}\right)^{1/3} \sim 10^{18} \left({E_{52}\over n_{\rm
    ext}}\right)^{1/3}\;{\rm~cm}\;.
\end{equation}
For comparison, the Sedov radius of a supernova that ejects a 10
$M_\odot$ envelope could reach $\sim 5$ pc or more.

In GRB sources, with jets which we believe to be highly relativistic,
the orientation of the jet axis with respect to our line of sight will
strongly affect the source's appearance
(\citealt{d1995},
 \citealt{dalal02},
 \citealt{gpkw02},
 \citealt{rr05},
 \citealt{2007RMxAC..27..140G}),
because radiation from jet material will be Doppler beamed in the
direction of motion. Attempts to understand the luminosity function of
GRBs may have to take into account the statistics of orientation,
collimation,and velocity of the jet, as well as the jet's intrinsic
radiation properties.\\

Although our proposed synthesis of GRB physical properties is highly
conjectural and far from unique, we hope it will provide a framework
for discussing the integrated properties of these objects. We conclude
the review by discussing how future observations, experiments and
theoretical studies should enhance our understanding of the physical
properties underlying GRBs.


\section{PROSPECTS}
\label{sec:prospects}

GRB studies, especially the afterglow-enabled studies of the last ten
years, remain a young field. The years ahead are thus sure to bring
astonishing discoveries as the capabilities and experience of
observers improve, theorists make more and stronger ties to physical
theory, and new and upgraded facilities open vistas.  In this section,
we summarize the instrumental capabilities and theoretical
opportunities for near-term progress in GRB research.


\subsection{Facilities}
\label{sub:prospects:facilities}

Table~\ref{tab:missions} summarizes recent and near-term future GRB
missions, planned and proposed.  \swift\ \citep{gehrels04} is now the
primary mission and has excellent prospects for continued operation,
with an orbit that will be stable until at least 2020.  \fermi\ and
\textit{AGILE}, also in continuing operations, are providing added
burst detections with simultaneous high energy ($>$100\,MeV) coverage
that promises to redefine the maximum energies and Lorentz factors
that GRB engines are capable of producing.

\begin{table}

\refstepcounter{table}
\label{tab:missions}

\hspace*{-1.5in}{Table 2~~Recent and future GRB Missions}

\vspace*{0.1in}

\hspace*{-1.5in}{\scriptsize
\begin{tabular}{l p{1.2in} l l p{1.0in} l}

Mission & 
Trigger energy range & 
FOV & 
Detector area & 
Other wavelengths & 
GRB rate (yr$^{-1}$) \\\hline

BATSE       & 20 keV--1.9 MeV (LAD)  & $4\pi$ sr
            & 2025 cm$^{2}$ per LAD &     & 300 \\

            & 10 keV--100 MeV (SD)  &
            & 127 cm$^{2}$ per SD  &          & \\ \hline

 \hete      & 6--400 keV            &  3 sr
            & 120 cm$^{2}$ & X-ray  &  \\ \hline

 \swift     & 15--150 keV           & 1.4 sr
            & 5200 cm$^{2}$ 
            & UV, Opt, X-ray 
            & 100 ($\sim$10\% SGRBs)\\ \hline

{\it AGILE} & 30 MeV--50 GeV & $\sim$3 sr & & Hard X-ray \\ \hline

{\it Fermi} & 20 MeV--300 GeV      & $>$2 sr (LAT)
            & $>$8000 cm$^{2}$ (LAT)   &   & 50  \\

            & 8 keV--1 MeV    & 9.5 sr (GBM)
            & 126 cm$^{2}$ (GBM-LED) &  & \\

            & 150 keV--30 MeV &
            & 126 cm$^{2}$ (GBM-HED) &  & \\ \hline

{\it SVOM} & 4 keV--300 keV & 2 sr (CXG)
              & & Optical, X-ray  & 80 \\ 

            & 50 keV--5 MeV & $89^{\circ}$ $\times$ $89^{\circ}$
              (GRM) & & \\ \hline

{\it JANUS} & 1--20 keV & 4 sr & & Near-IR  & 25 (high $z$) \\ \hline

{\it EXIST} & 5--600 keV & $\sim$3.6 sr (HET) & 5.96 m$^{2}$
               (HET) & Optical, near-IR,\newline X-ray & 300 \\\hline

\end{tabular}

\vspace{0.1in}

\hspace*{-1.5in}\parbox[t]{6.75in}{%
Note: LAD, Large Area Detector; SD, Spectroscopy Detector; LAT, Large
Area Telescope; LED, Low Energy Detector; HED, High Energy Detector;
CSG, X-ray/Gamma-ray Camera; GRM, Gamma-Ray Monitor; GBM, GLAST Burst
Monitor; SGRB, short GRB.}
}

\end{table}


The SVOM mission \citep{pwzb08}, currently under development, promises
4 to 300\,keV coverage and \swift-like slews that will bring
\xray\ and optical telescopes to bear on burst positions.  The
proposed \textit{JANUS} small explorer \citep{roming08} would focus on
high-redshift bursts, detecting prompt emission over the 1 to 20\,keV
band and slewing to observe afterglows with a near-infrared telescope
(50\,cm aperture, 0.7 to 1.7 $\mu$m coverage).  The proposed large
mission \textit{EXIST} has also added fast-response slews and focusing
\xray\ and near-infrared telescopes to its original complement of hard
\xray\ imaging detectors \citep{grindlay08}.


Current and near-future ground-based facilities include a fast-growing
array of robotic telescopes primed to respond to burst alerts, ongoing
improvements to the instrumentation and capabilities of large-aperture
telescopes, a new generation of air-Cerenkov TeV facilities, and with
the EVLA initiative, the most significant upgrade to a
high-sensitivity radio facility in decades.

Rapid follow-up of GRB discoveries is occurring both in space and on
the ground; the \swift\ UVOT observes $\sim$80 GRBs per year within 2
minutes, detecting about half of them.  On the ground, fast new
telescopes such as ROTSE-III \citep{akerlof03}, RAPTOR
\citep{Wozniak06}, Pi-of-the-sky \citep{burd05}, and REM \citep{Zerbi01}
are able to point at GRBs within $10-20$ s, and larger facilities like
the seven-band GROND \citep{gbc+08} respond on 10-minute timescales.
Ready availability of sensitive CCDs and HgCdTe detectors, combined
with rapid-slew mounts and autonomous software systems, should enable
further expansion of rapid-response telescopes with more larger
($D\simgt 2$\,m) facilities anticipated in the near future.

GRB-related programs continue to compete successfully for time on
premier optical facilities.  Large-aperture ($D\simgt 6$\,m)
telescopes provide the spectroscopic observations necessary for GRB
redshift measurements.  At the VLT, a rapid response mode has gone
further to provide time-sequence observations revealing variable
absorption from the host galaxy, and the impending commissioning of
the X-Shooter spectrograph \citep{kdh+09} will enable full UV to NIR
characterization of afterglow spectra with a single integration.  In
addition to the TOO opportunities provided by GRB alerts, multiple
host galaxy survey programs are underway.

At radio wavelengths, the workhorse facilities have been the VLA
(e.g., \citealt{ccf+08}) and WSRT (e.g., \citealt{vkw+07}), providing data
primarily in the $1-10$ GHz at sensitivities in the $0.2-1.0$ mJy
range; the Expanded VLA (EVLA) upgrade will improve the sensitivity of
that facility to $\sim$10\,\uJy.  Looking ahead, by 2012, the ALMA
array will be on line operating in the higher frequency range $90 -
950$ GHz with $>100$ times the sensitivity of the VLA.  The peak in
the synchrotron spectrum for a wide range of GRB parameters lies in
the ALMA range, and should make it a powerful future tool for radio
observations.

Observations of GRBs at TeV energies can be performed by both
narrow-field air Cerenkov facilities and wide-field water Cerenkov
detectors; the latter approach, at Milagro, yielded the tentative
detection of prompt TeV emission for GRB 970417A \citep{atkins03}.
There has been no detection to-date of prompt or afterglow emission
with the air Cerenkov facilities, but significant effort in pursuit of
burst alerts is underway at MAGIC, HESS, and VERITAS.  The MAGIC dish,
in particular, has a rapid-response mode that has provided
observations of multiple bursts within a minute of trigger
\citep{Albert06}.  Future facilties will seek to lower detection
thresholds to $E\sim 100$\,GeV, which would provide a
significantly-expanded horizon within which GRB sources will be
visible, rather than attenuated by photon-photon interactions.

In space, the upcoming \hst\ SM4 promises to revive that facility, not
only providing new and resuscitated instruments, but also the gyros
necessary for flexible and fast-response scheduling. In the future,
the capabilities of \textit{JWST} will provide an excellent resource
for high signal-to-noise observations of highly obscured and
high-redshift GRBs.

Within the \xray\ band, the \swift\ XRT has redefined all expectations
for the characterization of \xray\ afterglows, and at the same time,
proved a highly effective facility for refining the multi-arcmin
localizations provided by other GRB missions.  \chandra, \xmm, and
\suzaku\ continue to be active in GRB observations, and future
missions including the Indian ASTROSAT and Japan/US Astro-H promise
added capabilities in the near future.  Dramatic improvements will
await the next-generation GRB facilities (Table~\ref{tab:missions}) or
the arrival of the \textit{International X-ray Observatory}.


\subsection{Multimessenger Aspects}
\label{sub:prospects:multimess}

Given the rapid ongoing expansion of the capabilities of
non-electromagnetic detector facilities, and the extreme luminosity
and time-specificity of GRB sources, it will not be long before we
either have the first coincident, multi-messenger detection of a GRB,
or realize limits on the non-electromagnetic emissions of GRBs that
challenge our current understanding.  

The same shocks which are thought to accelerate electrons responsible
for the non-thermal $\gamma$-rays in GRBs should also accelerate
protons, leading ultimately to copious emission of high-energy
neutrinos (\citealt{waxman04}, \citealt{waxman04apj},
\citealt{waxman06}, \citealt{dermer05}).  The maximum proton energies
achievable in GRB shocks are $E_p\sim 10^{20}$\,eV, comparable to the
highest energies measured with large cosmic ray ground arrays
\citep{hayashida99}. For this, the acceleration time must be shorter
than both the radiation or adiabatic loss time and the escape time
from the acceleration region \citep{waxman95}. The accelerated protons
can interact with the fireball photons, leading to charged pions,
muons and neutrinos.  For internal shocks producing observed 1 MeV
photons this implies $\geq 10^{16}$ eV protons, and neutrinos with
$\sim 5\%$ of that energy, $\epsilon_\nu\geq 10^{14}$ eV
\citep{waxmanbahcall97}. Another copious source of target photons in
the UV is the afterglow reverse shock, for which the resonance
condition requires higher energy protons leading to neutrinos of
$10^{17}-10^{19}$ eV \citep{waxmanbahcall99}. Whereas photon-pion
interactions lead to higher energy neutrinos and provide a direct
probe of the shock proton acceleration as well as of the photon
density, inelastic proton-neutron collisions may occur even in the
absence of shocks, leading to charged pions and neutrinos
\citep{derishev99} with lower energies than those from photon-pion
interactions. The typical neutrino energies are in the $\sim$ 1-10 GeV
range, which could be detectable in coincidence with observed
GRBs. This is the province of projects like AMANDA, IceCube and
ANTARES. Neutrino astronomy has the advantage that we can see the
universe up to $\sim$~EeV energies. By contrast, the universe becomes
opaque to $\gamma$-rays above $\sim$ TeV energies through absorption
by the infrared background.

The last and most challenging frontier is that of gravitational
radiation, which is largely unknown territory. A time-integrated
luminosity of the order of a solar rest mass ($\sim 10^{54}$ erg) is
predicted from progenitor models involving merging compact objects,
while that from collapsar models is less certain, and expected to be
lower by at least one order of magnitude.  

Ground-based facilities, like LIGO, TAMA and VIRGO, are currently
seeking the first detection of these stellar-scale, high-frequency
($\nu\simgt 50$\,Hz) sources. The observation of associated
gravitational waves would be facilitated if the mergers involve
observed short GRB sources; and conversely, it may be possible to
strengthen the case for (or against) \mbox{NS-NS} or \mbox{NS-BH}
progenitors of short bursts if gravitational waves were detected (or
not) in coincidence with some events. The technical challenge of
achieving the sensitivities necessary to measure waves from assured
sources should not be understated; neither, however, should the
potential rewards.

The Enhanced LIGO interferometers will be on line in 2009 with ability
to detect NS binary mergers to 20\,Mpc.  The Advanced LIGO
interferometers on line in 2014 will extend the distance to 200\,Mpc.
Short GRBs, if produced by mergers as is thought to be the case,
predict a cosmological rate density of $>10$ Gpc$^{-1}$ yr$^{-1}$ with
likely rate of $\sim$300 Gpc$^{-1}$ yr$^{-1}$
(\citealt{2005ApJ_633_1076O}, \citealt{2007ApJ_667_1048O},
\citealt{2008ApJ_672_479O}, \citealt{2008ApJ_675_566O},
\citealt{Nakar07}).  This translates into an Advanced LIGO detection
rate $\sim$10 yr$^{-1}$.  The density estimates include mergers with
$\gamma-$ray jets not aimed in our direction, so not all LIGO
detection would be in coincidence with GRBs.  LIGO is already
providing useful upper limits, as with GRB\,070201 described in
Section III.


\subsection{Cosmology}
\label{sub:prospects:cosmo}

\begin{figure}
\begin{center}
\includegraphics[scale=0.8]{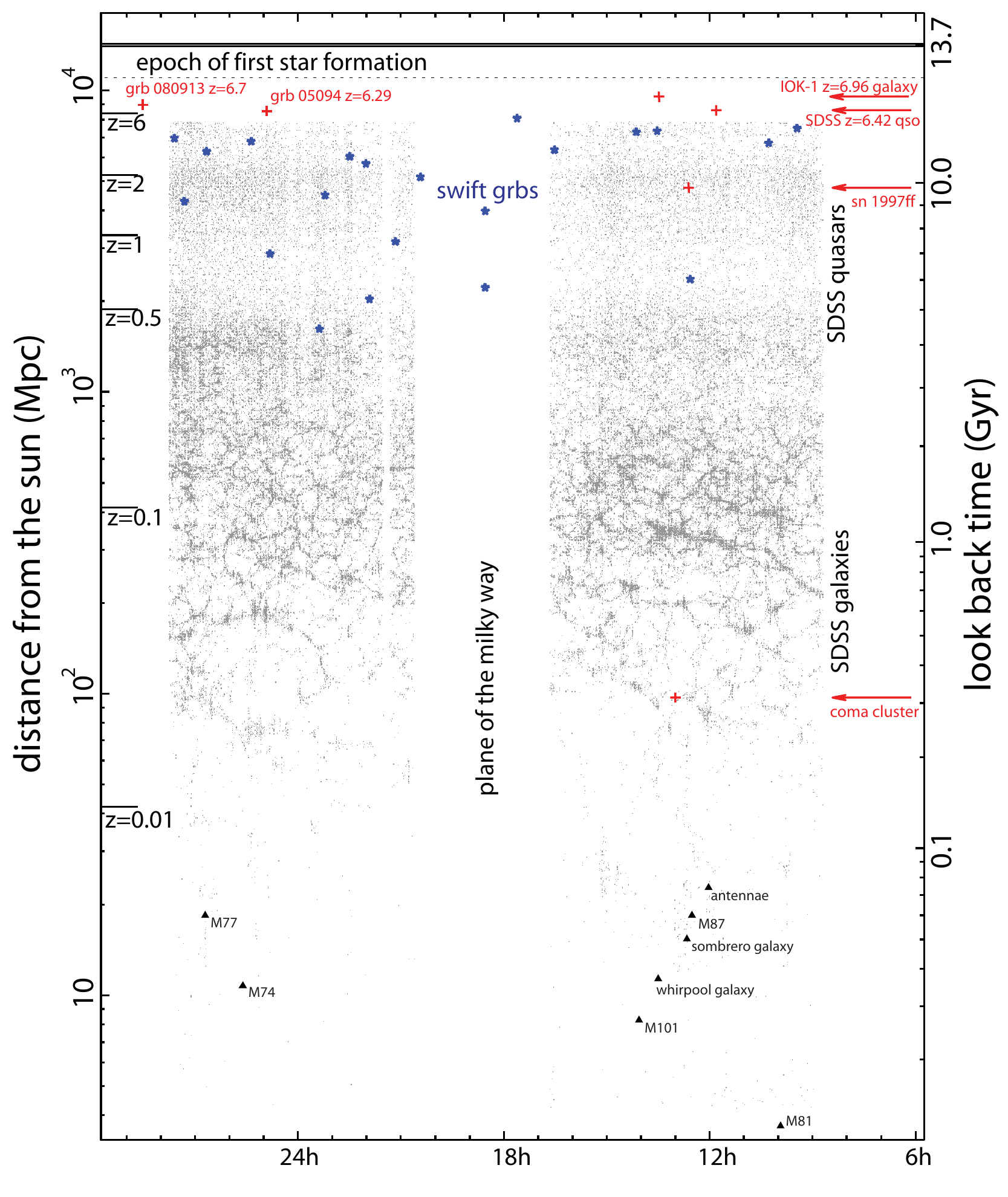}
\end{center}
\caption{\small
  A 360$^{\circ}$ vista showing the entire sky, with visible
  structures stretching back in distance, time and redshift.  The most
  distant light we observe comes from the radiation left over from the
  Big Bang: the CMB. As we descend the chart, we find the most distant
  objects known, followed by a web of Sloan Digital Sky Survey (SDSS)
  quasars and galaxies. Closer to home, we start to see a collection
  of familiar ``near'' galaxies (triangles).  Also marked are all
  Swift GRBs with known distances (blue stars); SN 1997ff, the most
  distant type Ia supernova at $z=1.7$; and the archetypal large
  galaxy cluster, the Coma cluster. The redshift distances of most
  distant GRBs are comparable to the most distant galaxies and quasars
  (adapted from \citealt{errnandv06}.)  }
\label{figXV}
\end{figure}

One of the frontiers of modern cosmology lies at high redshift,
$z\simgt 6$, when the first non-linearities developed into
gravitationally-bound systems, whose internal evolution gives rise to
stars, galaxies, and quasars; and when the light emitted from these
first collapsed structures diffuse outwards to reionize the Universe.
As the (temporarily) brightest source of photons in the cosmos (see
Figure~\ref{figXV}), the demise of these first generations of massive
stars in GRB explosions defines the challenge of elucidating the end
of the cosmic ``dark ages'' (\citealt{2000ApJ_536_1L},
\citealt{2002ApJ_575_111B}).

Apart from revealing a site of high-redshift star-formation, each such
high-redshift burst has the potential to help constrain local element
abundances in its host galaxy, information that will be impossible to
gather by other means until the advent of $D\simgt 20$\,m telescopes.
Even more exciting, each burst has the potential to reveal the extent
of intergalactic reionization, at its redshift and along that line of
sight (\citealt{miralda98}, 
       \citealt{mlz+08}, 
       \citealt{mf08}).

At the highest redshifts, $z\simgt 10$, there is growing theoretical
evidence indicating that the first luminous objects to form were very
massive stars $M > 100 M_\odot$.  Depending on whether these stars
retain their high masses until death, and whether a fast-rotating core
is prerequisite to the GRB phenomenon, these Pop~III stars might
provide the progenitors for the most luminous, highest-redshift GRBs
\citep{2003ApJ_591_288H}.

The first GRBs and supernovae may also be important for another
reason: they may generate the first cosmic magnetic fields. Mass loss
(e.g. via winds) would disperse magnetic flux along with the heavy
elements. The ubiquity of heavy elements in the Lyman alpha forest
indicates that there has been widespread diffusion from the sites of
these early supernovae, and the magnetic flux could have diffused in
the same way. This flux, stretched and sheared by bulk motions, could
be the ``seed'' for the later amplification processes that generate
the larger-scale fields pervading disc galaxies.


\subsection{Theoretical Prospects}
\label{sub:propects:theory}
 
What can we expect in the wave of matching theoretical progress?  This
is more difficult to disucss because theory often develops on a
shorter timescale than observations and experiments, and so we cannot
forsee most future developments. Although some of the features now
observed in GRB sources (especially afterglows) were anticipated by
theoretical discussions, the recent burst of observational discovery
has left theory lagging behind.  There are, however, some topics on
which we do believe that there will be steady work of direct relevance
to interpreting observations.

One of the most important is the development and use of hydrodynamical
codes for numerical simulation of GRB sources with detailed physics
input. Existing two and three dimensional codes have already uncovered
some gas-dynamical properties of relativistic flows unanticipated by
analytical models \citep{mb08} but there are some key questions that
they cannot yet address. In particular, higher resolution is needed
because even a tiny mass fraction of baryons loading down the outflow
severely limits the maximum attainable Lorentz factor.  What is more,
jets are undoutedly susceptible to hydrostatic and hydromagnetic
turbulence. We must wait for useful and affordable three dimensional
simulations before we can understand the nonlinear development of
instabilities. Well-resolved three dimensional simulations are
becoming increasingly common and they rarely fail to surprise us.  The
symmetry-breaking involved in transitioning from two to three
dimensions is crucial and can lead to qualitatively new phenomena. A
particularly important aspect of this would be to link in a
self-consistent manner the flow within the accretion disk to that in
the outflowing gas, allowing for feedback between the two
components. A self-consistent model incorporating inflow and outflow
must also explain how some fraction of the material can acquire more
than its share of energy (i.e., a high enthalpy or $p/\rho$).

Particle acceleration and cooling is another problem that seems ripe
for a more sophisticated treatment.  Few would dispute the statement
that the photons which bring us all our information about the nature
of GRBs are the result of particle acceleration in relativistic shocks
(\citealt{2008arXiv0810.4014M}, \citealt{2008ApJ...673L..39S},
\citealt{2007ApJ_671_1877R}, \citealt{2007ApJ_655_375K}) or turbulence
(\citealt{2007arXiv0706_1818G}, \citealt{2009ApJ_692L_40Z},
\citealt{2008ApJ...688..462C}).  Since charged particles radiate only
when accelerated, one must attempt to deduce from the spectrum {\it
  how} the particles are being accelerated, {\it why} they are being
accelerated, and to identify the macroscopic source driving the
microphysical acceleration process.

Collisionless shocks are among the main agents for accelerating ions
as well as electrons to high energies whenever sufficient time is
available (e.g., \citealt{blandford87},
\citealt{achterberg01}). Particles reflected from the shock and from
scattering centres behind it in the turbulent compressed region have a
good chance of experiencing multiple scattering and acceleration by
first-order Fermi acceleration when coming back across the shock into
the turbulent upstream region. Second-order or stochastic Fermi
acceleration in the broadband turbulence downstream of collisionless
shocks will also contribute to acceleration. In addition, ions may be
trapped at perpendicular shocks. The trapping is a consequence of the
shock and the Lorentz force exerted on the particle by the magnetic
and electric fields in the upstream region. With each reflection at
the shock the particles gyrate parallel to the motional electric
field, picking up energy and surfing along the shock surface. All
these mechanisms are still under investigation, but there is evidence
that shocks play a most important role in the acceleration of cosmic
rays and other particles to very high energies.

Another topic on which further work seems practicable concerns the
kinematics of ultrarelativistic jets \citep{2007RMxAC..27..140G}.
While it seems probable that we are using the correct ingredients of
special relativity and a collimated outflow, it is equally true that
no detailed model yet commands majority of support. We can still
expect some surprises from studies related to the appearence of
relativistic shocks in unsteady jets.

The most interesting problem remains, however, in the nature of the
central engine and the means of extracting power in a useful
collimated form. In all observed cases of relativistic jets, the
central object is compact, either a neutron star or black hole, and is
accreting matter and angular momentum. In addition, in most systems
there is direct or indirect evidence that magnetic fields are present
-- detected in the synchrotron radiation in galactic and extragalactic
radio sources or inferred in collapsing supernova cores from the
association of remnants with radio pulsars. This combination of
magnetic field and rotation may be very relevant to the production of
relativistic jets.

\section{CONCLUSIONS}
\label{sec:conclude}
 
\begin{figure}
\begin{center}
\includegraphics[scale=0.5]{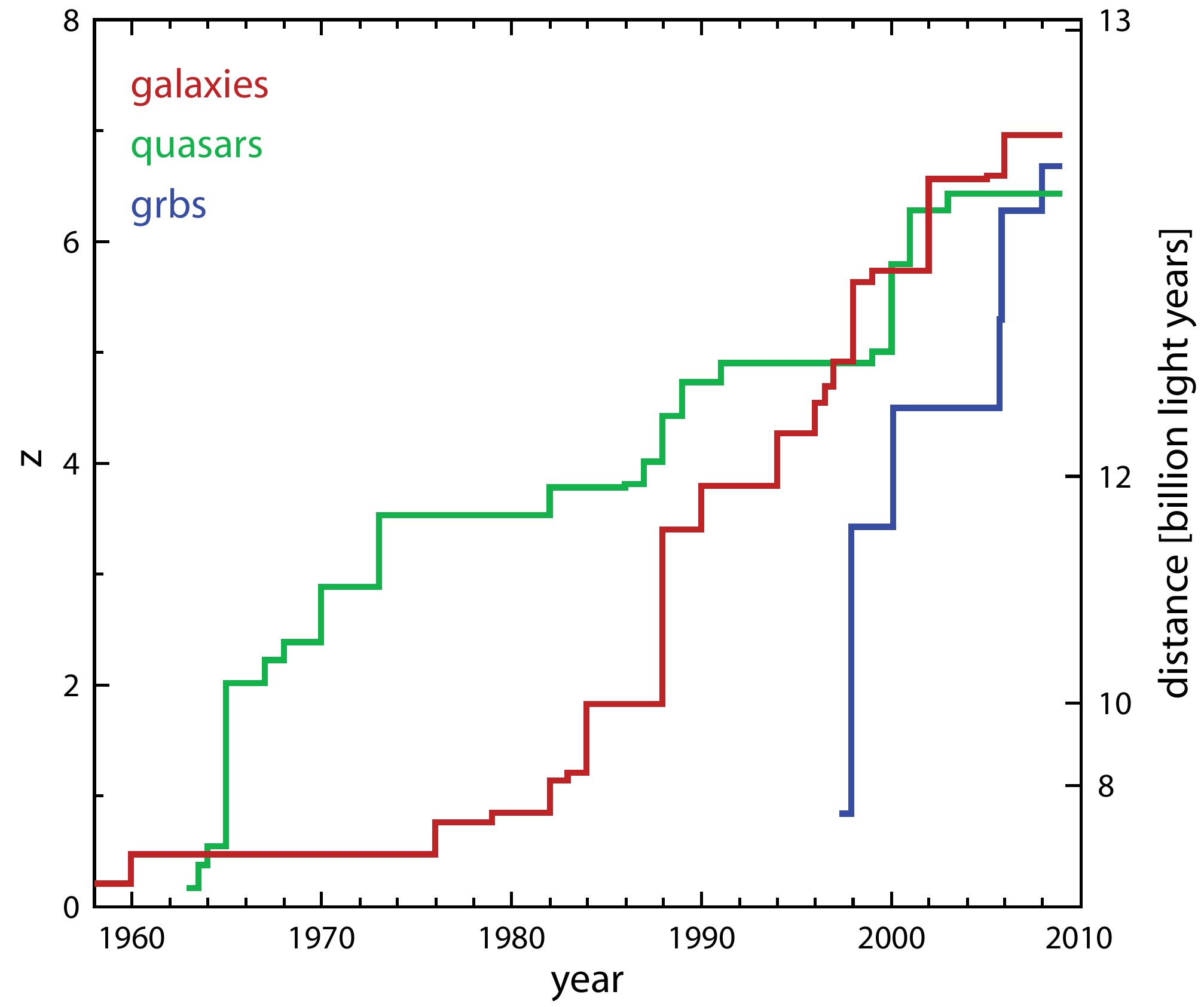}
\end{center}
\caption{\small
   High-z record holders.  The history of most distant objects
  detected in categories of galaxies, quasars, and GRBs
  (R. McMahon, N. Tanvir, priv. comm.). }
\label{figVb}
\end{figure}

Thanks primarily to the burst discoveries and observations of the
\swift\ satellite, the last five years have been tremendously
productive ones for GRB research.  The identification of short burst
afterglows has confirmed long-held suspicions that GRBs have at least
two fundamentally different types of progenitors; subsequent studies
of short burst afterglows and host galaxies have furnished valuable
information on the nature of their progenitors and provided hopeful
indications for next-generation gravity-wave observatories.  The
discovery of the first two bursts at $z>6$, before sources reionized
most of the hydrogen in the universe, has proven the value of GRBs as
probes of the earliest cosmic epochs and and extended GRB observations
beyond the redshifts of the most distant known quasars (see
Figure~\ref{figVb}).  A flood of prompt burst alerts has fed the
queues of more than a dozen dedicated robotic telescopes, and prompted
fast-response multi-epoch high-resolution spectroscopy from the
largest telescopes.  Overall, bright GRBs continue to attract the
attention of astronomers of all types, with premier facilities across
the electromagnetic spectrum poised to respond to the next spectacular
event.  In the near future, we hope that gravity-wave and high-energy
neutrino astronomers will be rewarded for their decades of persistent
effort.

The complexity -- not to mention sheer volume -- of data in the
\swift\ era have inevitably raised challenges to the prior
interpretation of bursts and their afterglows.  However, careful
consideration of the biases inherent to the \swift\ mission observing
strategies, along with the large number of events now available for
analysis, are gradually enabling the construction of a single coherent
picture, via a multidisciplinary approach that addresses data from
across the electromagnetic spectrum. It is one of the challenges of
contemporary research to infer the underlying physical structure of
GRBs in the belief that this is simple, despite the complex character
of the observations.

GRBs provide us with an exciting opportunity to study new regimes of
physics. As we have described, our rationalization of the principal
physical considerations combines some generally accepted features with
some more speculative and controversial ingredients. When confronted
with observations, it seems to accommodate their gross features but
fails to provide us with a fully predictive theory. What is more
valuable, though considerably harder to achieve, is to refine models
like the ones advocated here to the point of making quantitative
predictions, and to assemble, assess and interpret observations so as
to constrain and refute these theories.  What we can hope of our
present understanding is that it will assist us in this endeavour.

There are good prospects for a continued high rate of discovery going
forward.  \swift\ is likely to remain operational for many years and
{\it Fermi} and {\it AGILE}, now on orbit, are providing new insights
at higher energies.  Instruments currently in development or planning
have the potential to enable further qualitative advances.  GRBs are
among the most extraordinary of astronomical phenomena and will, with
our present and future capabilities, continue providing a unique
window into the extreme reaches of the Universe.


\section*{Acknowledgements}

We are indebted to many colleagues for their contributions to this
review.  We are very grateful to John Cannizzo for his insightful
comments and tireless help with editing the manuscript.  Several
colleagues provided figures and data that we much appreciate.  They
are Edo Berger, Andy Fruchter, Alexander Kahn, Yuki Kaneko, David
Morris, Kim Page, Mario Juric, Jacob Palmatier, Bryan Penprase,
J. X. Prochaska, Judy Racusin, Taka Sakamoto, Richard McMahon, and
Nial Tanvir.  Our views on the topics discussed here have been
clarified through discussions with Josh Bloom, Jonathan Granot,
William Lee, Peter Meszaros, Udi Nakar, Tsvi Piran, X. Prochaska,
Chryssa Kouveliotou, Pawan Kumar, Edo Berger, Alicia Soderberg, Brad
Cenko, Dale Frail, Jane Charlton, David Burrows, Doug Cowen, Richard
O'Shaughnessy, Kris Belczynski, Kris Stanek, Matthew McQuinn,
Hsiao-Wen Chen, Andrew MacFadyen, Sam Oates, Martin J. Rees, Stephan
Rosswog, Eli Waxman, Stan Woosley, Chris Fryer, Guillermo
Garcia-Segura, Norbert Langer, Sung-Chul Yoon, Dan Kasen, and Annalisa
Celotti.  We thank J. D. Myers and Rion Parsons for assistance with
graphics.  Finally, we thank the 200 members of the \swift\ team for
the wonderful new findings reported here.

This work is partially supported by NSF: PHY-0503584 (ER-R), 
DOE SciDAC: DE-FC02-01ER41176 (ER-R).







\end{document}